\documentclass[sigconf]{acmart}

\renewcommand\footnotetextcopyrightpermission[1]{} 
\setcopyright{none}

\usepackage{booktabs}   
\usepackage{subcaption} 

\usepackage{graphicx}
\usepackage{float}
\usepackage{stfloats}
\usepackage{algorithm}
\usepackage{algpseudocode}
\usepackage{subcaption}
\usepackage{multirow}
\usepackage{makecell}
\usepackage{wrapfig}
\usepackage{xcolor,lipsum,subcaption}
\usepackage{enumitem}
\usepackage{listings}
\usepackage{minted}
\usepackage{microtype}
\usepackage{makecell}

\begin{document}

\newcommand{\toolName}{FlashQuant}
\newcommand{\formatName}{Tile-COO}

\title[{\toolName}]{{\toolName}: Sparse-Dense Fusion for Memory-Efficient Outlier-Aware LLM Inference} 

\author{Junqing Lin}
\email{linjunqing@mail.ustc.edu.cn}
\orcid{0009-0007-1455-8725}
\affiliation{
    \institution{University of Science and Technology of China}
    \city{Hefei}
    \country{China}
}
\author{Jingwei Sun}
\authornote{Co-corresponding author}
\email{sunjw@ustc.edu.cn}
\orcid{0000-0001-5098-1503}
\affiliation{
    \institution{University of Science and Technology of China}
    \city{Hefei}
    \country{China}
}

\author{Zhengding Hu}
\email{huzd@mail.ustc.edu.cn}
\affiliation{
    \institution{University of Science and Technology of China}
    \city{Hefei}
    \country{China}
}

\author{Guangzhong Sun}
\orcid{0000-0002-0794-7681}
\email{gzsun@ustc.edu.cn}
\affiliation{
    \institution{University of Science and Technology of China}
    \city{Hefei}
    \country{China}
}


\makeatletter
\let\old@fnsymbol\@fnsymbol
\def\@fnsymbol#1{%
	\ifnum#1=1\relax
	\textdagger
	\else
	\old@fnsymbol{#1}%
	\fi
}
\makeatother

\begin{abstract}
Low-bit quantization reduces the memory footprint and computational cost of large language model (LLM) inference. However, high-magnitude outlier weights can induce substantial quantization errors and degrade model accuracy. Outlier-aware quantization addresses this issue by retaining outliers in high precision while quantizing the remaining weights, resulting in a low-bit dense GEMM path and a high-precision sparse SpMM path. Existing implementations execute these paths in separate GPU kernels, despite their shared activations and outputs, thereby missing opportunities for intra-operator reuse and incurring redundant global-memory accesses. This inefficiency is particularly pronounced in memory-bound decoding workloads.
We propose FlashQuant, a content-sharing execution framework for outlier-aware W4A16 decoding. FlashQuant fuses the dense GEMM and sparse outlier SpMM paths into a single GPU kernel, enabling on-chip reuse of activation and output tiles across heterogeneous computations. It introduces three key techniques: sparse-dense tiling, which aligns outlier processing with dense GEMM tiles; Tile-COO outlier encoding, which enables efficient sparse access and reduces shared-memory bank conflicts; and pipelined scheduling, which overlaps computation with data movement. Experiments show that FlashQuant reduces outlier-processing overhead, achieving 2.74×–4.18× speedup over cuBLAS BF16 and up to 1.53× speedup over the strongest unfused outlier-aware baseline.

\end{abstract}


\maketitle

\section{Introduction}
Large language models (LLMs) deliver strong performance but impose substantial inference-time computational and memory costs~\cite{team2024qwen2,yang2025qwen3,grattafiori2024llama}. As parameter counts continue to grow, efficient deployment becomes increasingly difficult, particularly in latency-sensitive or resource-constrained environments. For example, Qwen3-235B~\cite{yang2025qwen3} requires more than 470~GB of memory in BF16 precision, which exceeds the memory capacity of widely used GPUs such as the NVIDIA A100 Tensor Core GPU.

Weight-only quantization~\cite{lin_awq_2025,dettmers_spqr_2023,dettmers_llmint8_nodate,shao_omniquant_2024,frantar_gptq_2023,guan_aptq_2024,kim_squeezellm_2024,ashkboos_quarot_nodate,shang_pb_llm_2023,li_llm-mq_nodate,li_fast_2024,huang_slim-llm_2024,xiao2023smoothquant,yuan2023rptq} is widely adopted to reduce the memory footprint and bandwidth cost of LLM inference. By storing weights in low-bit formats, such as INT4, while retaining activations in higher precision, it offers a favorable accuracy--efficiency trade-off for decoding workloads. However, aggressive low-bit quantization remains vulnerable to high-magnitude outlier weights: quantizing these outliers together with regular weights can cause substantial reconstruction errors and degrade model accuracy.

Outlier-aware quantization~\cite{kim_squeezellm_2024,dettmers_llmint8_nodate,huang_slim-llm_2024,li_llm-mq_nodate,shang_pb_llm_2023} mitigates quantization error by preserving a small fraction of outlier weights in high precision while quantizing the remaining weights. For example, retaining 1\% of outlier weights during INT4 quantization of LLaMA3-8B reduces WikiText2 perplexity \textbf{from 6.896 to 5.979}. This strategy decomposes the original matrix multiplication into two computational paths:
\begin{equation}
Y = X \cdot W \approx X \cdot \text{dequant}(W_Q) + X \cdot W_O,
\end{equation}
where $W_Q$ is the low-bit dense weight matrix and $W_O$ is the sparse high-precision outlier matrix. The dense component is computed via low-bit GEMM, while the outlier component is typically handled by high-precision SpMM. Although this formulation improves quantization accuracy, it introduces a mixed dense–sparse execution pattern into the inference pipeline.

\begin{figure}[h]
    \centering
    \includegraphics[width=0.95\linewidth]{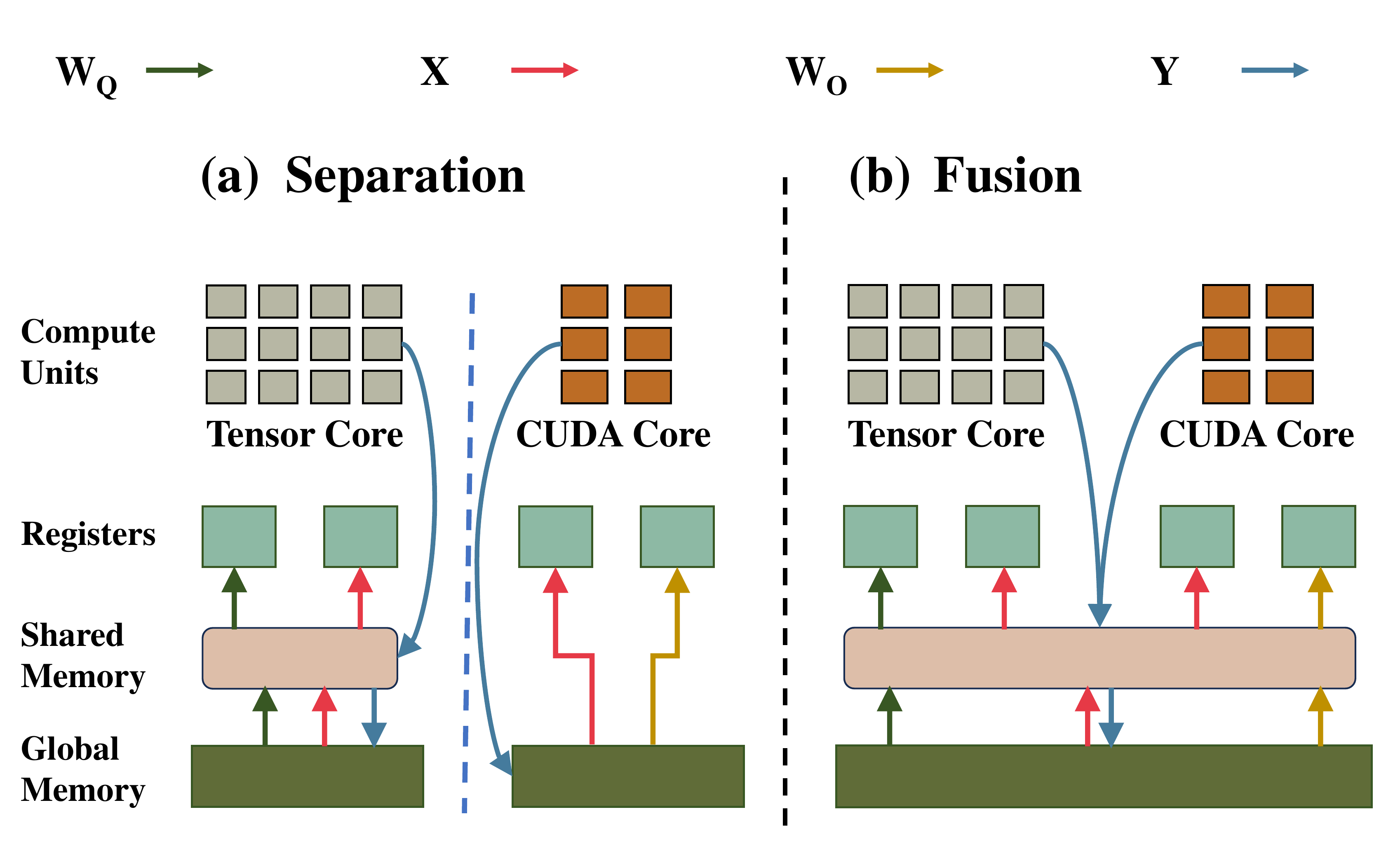}
    \caption{Execution flow of outlier-aware quantization kernels: (a) existing approaches; (b) {\toolName}.}
    \label{fig:motivation}
\end{figure}

Existing outlier-aware quantization systems typically execute the dense and sparse paths with separate GPU kernels, as illustrated in Figure~\ref{fig:motivation}(a). Although this design leverages highly optimized dense GEMM and sparse SpMM libraries, it overlooks a key data-sharing opportunity: both paths consume the same activation matrix and accumulate into the same output matrix. Separate execution therefore reloads activation tiles from global memory and writes partial outputs independently, introducing redundant memory traffic. This overhead is especially pronounced in memory-bound decoding workloads, where even a small fraction of preserved outliers can cause disproportionate latency.

To address this inefficiency, we propose {\toolName}, a unified execution framework for outlier-aware W4A16 decoding. {\toolName} fuses the low-bit dense GEMM path and the high-precision sparse outlier path into a single GPU kernel. Unlike conventional kernel fusion, which mainly combines producer--consumer operators or merges independent kernels to reduce launch overhead, {\toolName} introduces content-sharing fusion: two heterogeneous computation paths derived from the same mathematical operator are co-scheduled over a common tiled execution space. This design enables on-chip reuse of shared data across dense and sparse computations, thereby eliminating redundant global-memory accesses.
Realizing this execution model is challenging because dense GEMM and sparse SpMM exhibit fundamentally different execution patterns. Dense GEMM relies on regular, hardware-aligned tiling and Tensor Core execution, whereas sparse outlier processing requires irregular metadata accesses, nonuniform nonzero handling, and CUDA-core accumulation. {\toolName} addresses these challenges with three co-designed techniques. First, sparse--dense tiling maps outlier computation onto the dense GEMM tile hierarchy, establishing a common schedule for both paths. Second, {\formatName} organizes outliers by tile boundaries and bucket assignments, enabling vectorized decoding and conflict-free sparse accumulation. Third, a pipelined execution strategy overlaps data movement, dequantization, outlier decoding, and computation to hide memory latency.

Experiments on NVIDIA GeForce RTX 3090, 4090, and 5090 GPUs show that {\toolName} effectively mitigates the performance overhead of outliers, achieving a $2.74\times$--$4.18\times$ quantization speedup. It also outperforms the strongest existing outlier-aware implementation, which combines state-of-the-art dense quantized and sparse operators~\cite{frantar2025marlin,gale_sparse_2020}, by up to $1.53\times$.

In summary, this paper makes the following contributions:
\begin{itemize}
    \item We identify redundant memory access as a key system bottleneck in outlier-aware quantized inference. Existing implementations execute the quantized dense path and sparse outlier path in separate GPU kernels, limiting inference performance.
    \item We propose {\toolName}, a unified sparse-dense execution framework that fuses low-bit GEMM and high-precision SpMM into a single GPU kernel. {\toolName} further improves efficiency through sparse-dense tiling, {\formatName}, and pipelined scheduling.
    \item We evaluate {\toolName} on representative LLM operators across multiple GPU generations, showing consistently lower outlier-processing overhead and throughput improvements over outlier-aware baselines.
\end{itemize}

\section{Background}
\subsection{LLM Inference}
LLMs are predominantly based on the transformer architecture~\cite{zhao2023survey}, which consists of stacked attention and feedforward layers. Each layer performs multiple GEMM operations, including query-key-value (QKV) projections, the self-attention output projection, and the two linear transformations in the feedforward network. In typical inference workloads, these GEMM kernels dominate both total FLOPs and end-to-end latency~\cite{xia2023flash}. Therefore, improving GEMM efficiency is critical for accelerating practical LLM inference.

LLM inference is typically divided into two phases: \textit{prefilling} and \textit{decoding}. During prefilling, the model processes the entire input prompt in a single forward pass, enabling all tokens to be propagated through the network in parallel. In contrast, decoding proceeds autoregressively, generating one token at a time conditioned on previously generated tokens. At each decoding step, the activation matrix $X \in \mathbb{R}^{B \times K}$ is multiplied by the projection weight matrix $W \in \mathbb{R}^{K \times N}$ to produce the next hidden states. Because the batch size $B$ is typically small in resource-constrained settings, decoding offers limited parallelism and increasingly becomes memory-bound~\cite{hong2024flashdecoding++}.

\subsection{Outlier-Aware Quantization}

Low-bit quantization is an effective technique for reducing the memory
footprint and computational cost of LLM inference~\cite{frantar_gptq_2023,huang_billm_2024}.
In a standard $\ b$-bit affine uniform quantization scheme, each
floating-point weight $w \in \mathbb{R}$ is mapped to an integer level
$q$ using a scale factor $s$ and an integer zero-point $z$. The
quantization and dequantization operations are defined as
\begin{equation}
    \label{eq:uniform_quant}
    q = \operatorname{round}\!\left(\frac{w}{s}\right) + z,
\end{equation}
\begin{equation}
    \label{eq:dequant}
    \hat{w} = s \cdot (q - z),
\end{equation}
where $ s = \frac{w_{\max} - w_{\min}}{2^b - 1},$ and $  z = \operatorname{round}\!\left(-\frac{w_{\min}}{s}\right)$.

However, aggressive quantization can substantially degrade model accuracy, particularly in layers whose weights have large dynamic ranges. Outlier-aware quantization~\cite{frantar_gptq_2023,dettmers_spqr_2023,kim_squeezellm_2024,shang_pb_llm_2023} mitigates this degradation by identifying high-magnitude weights, or \emph{outliers}, and storing them in higher precision, such as BF16, while quantizing the remaining weights using Equation~\eqref{eq:uniform_quant}. For example, preserving 1\% of outliers during INT4 quantization of LLaMA3-8B reduces WikiText2 perplexity from 6.896 to 5.979.

During inference, an outlier-aware operator is typically decomposed into two computational paths: GEMM for low-bit weights and SpMM for high-precision outliers. When executed as separate kernels, these two paths reload the same activation tile $X$ and write independent partial outputs to global memory. This separation prevents shared tiling, activation reuse, and coordinated scheduling between dense and sparse computations. In bandwidth-limited LLM decoding, the resulting redundant memory traffic becomes a major performance bottleneck. As a result, although the SpMM path accounts for only a small fraction of the weights, it can contribute disproportionately to end-to-end latency.

\subsection{GPU Architecture}

On GPUs, execution is organized into \emph{cooperative thread arrays} (CTAs), each scheduled on a single \emph{streaming multiprocessor} (SM). Within a CTA, threads are grouped into 32-thread \emph{warps}, the basic unit of \emph{single-instruction multiple-thread} (SIMT) execution. This hierarchical execution model exposes massive parallelism and is particularly effective for throughput-oriented workloads.

GPUs further rely on a hierarchical memory system that balances bandwidth, latency, and capacity. The main levels include high-capacity but high-latency \emph{global memory}, low-latency on-chip \emph{shared memory} visible to threads within a CTA, and per-thread \emph{registers}, which provide the lowest-latency storage. Shared memory is divided into multiple banks to enable concurrent warp-level accesses. However, when multiple threads access the same bank within a transaction, \emph{bank conflicts} occur, causing serialization and reducing shared-memory throughput.

In outlier-aware quantization, irregular memory access patterns often arise, and naive concurrent execution of quantized and outlier paths can incur redundant memory traffic. Achieving high performance therefore requires memory access strategies that exploit the GPU execution model and memory hierarchy to improve parallel efficiency while reducing access overhead.

\section{\toolName}

Outlier-aware quantization preserves model accuracy by retaining high-magnitude weights in higher precision. However, this benefit often incurs nontrivial performance overhead, as inference must execute an additional SpMM operation alongside the low-bit dense GEMM. This extra computation further aggravates the compute–memory imbalance on modern GPUs. For example, the NVIDIA GeForce RTX 5090 delivers up to 209.5 TFLOPs of FP16/BF16 throughput but only 1792 GB/s of memory bandwidth, requiring an arithmetic intensity exceeding 100 FLOPs/byte to fully utilize its compute units. In LLM inference, especially during autoregressive decoding, small batch sizes limit arithmetic intensity, making execution predominantly memory-bound. Consequently, even modest increases in memory traffic can directly degrade end-to-end latency and throughput.

Existing implementations typically handle outliers using generic SpMM kernels optimized for standalone sparse workloads. Each SpMM invocation therefore reloads the activation matrix from global memory and writes intermediate results separately, leading to duplicated memory accesses. Notably, GEMM and SpMM operate on the same input and output matrices, exposing substantial opportunities for intra-operator data reuse. A fused execution model could share activation tiles loaded into on-chip memory across both computation paths and accumulate partial results before write-back. Such fusion can reduce memory traffic and kernel launch overhead, thereby improving inference efficiency.

However, fusing GEMM with SpMM remains challenging because the two computation paths exhibit fundamentally different execution patterns. \textbf{First}, they impose different layout and tiling requirements: GEMM favors fixed, hardware-aligned tiles, whereas sparse outliers are often represented in irregular formats such as CSR, limiting on-chip reuse under naive fusion. \textbf{Second}, they operate at different granularities: GEMM leverages regular warp-level Tensor Core operations, while sparse outlier processing requires fine-grained accumulation on CUDA cores, complicating coordinated scheduling within a single kernel. \textbf{Third}, the nonuniform distribution of outliers leads to irregular activation accesses and imbalanced workloads, causing uncoalesced memory transactions, shared-memory bank conflicts, and load imbalance.

\begin{figure}[h]
    \centering
    \includegraphics[width=0.9\linewidth]{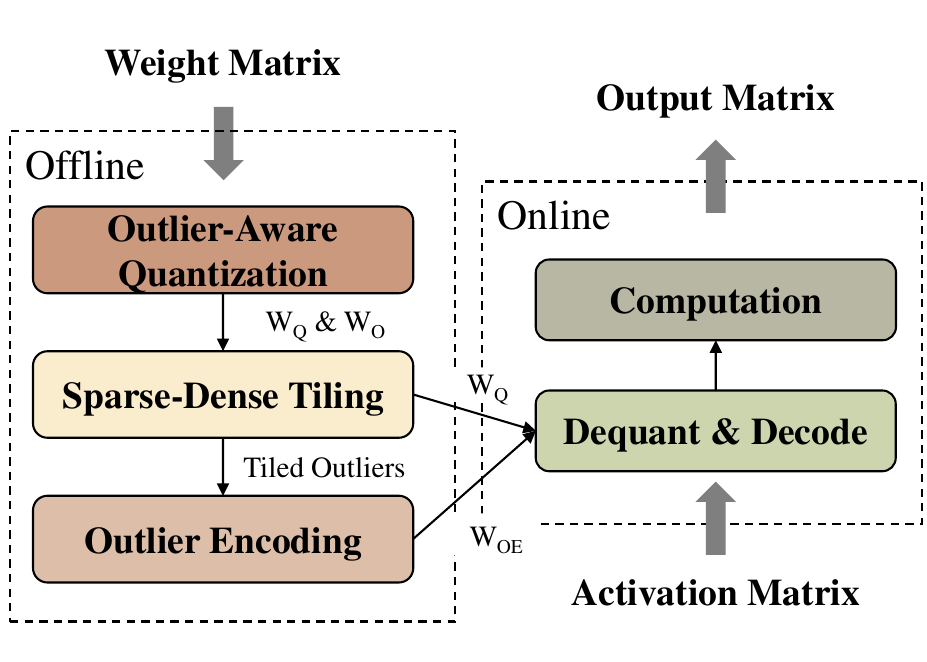}
    \caption{Overview of {\toolName}.}
    \label{fig:overview}
\end{figure}

To address these challenges, we present {\toolName}, a framework that unifies low-bit GEMM and high-precision SpMM via tile-level scheduling and optimized outlier encoding. {\toolName} targets the decoding stage of LLM inference, where batch sizes are typically small. As shown in Figure~\ref{fig:overview}, the framework consists of two stages: offline preprocessing and online inference.
During preprocessing, {\toolName} performs outlier-aware quantization, decomposing the original weight matrix into a low-bit quantized matrix \(W_Q\) and a sparse outlier matrix \(W_O\). It then applies a unified sparse--dense tiling scheme that partitions \(W_Q\) and \(W_O\) into aligned tiles, enabling the two computation paths to cooperatively load and reuse activations in shared memory. To further improve efficiency, {\toolName} encodes outliers in {\formatName} and reorders nonzero elements to improve load balance and reduce shared-memory bank conflicts.
During inference, {\toolName} fuses dense and sparse computation into a single kernel. Activations, quantized weights, and outlier tiles are first loaded into shared memory. The GEMM path performs low-bit matrix multiplication on Tensor Cores, while the SpMM path accumulates outlier contributions on CUDA Cores. By avoiding redundant activation reloads and intermediate global-memory write-backs, this fused design substantially reduces global memory traffic compared with separate GEMM and SpMM kernels.

\begin{figure*}[t]
    \centering
    \includegraphics[width=0.8\textwidth]{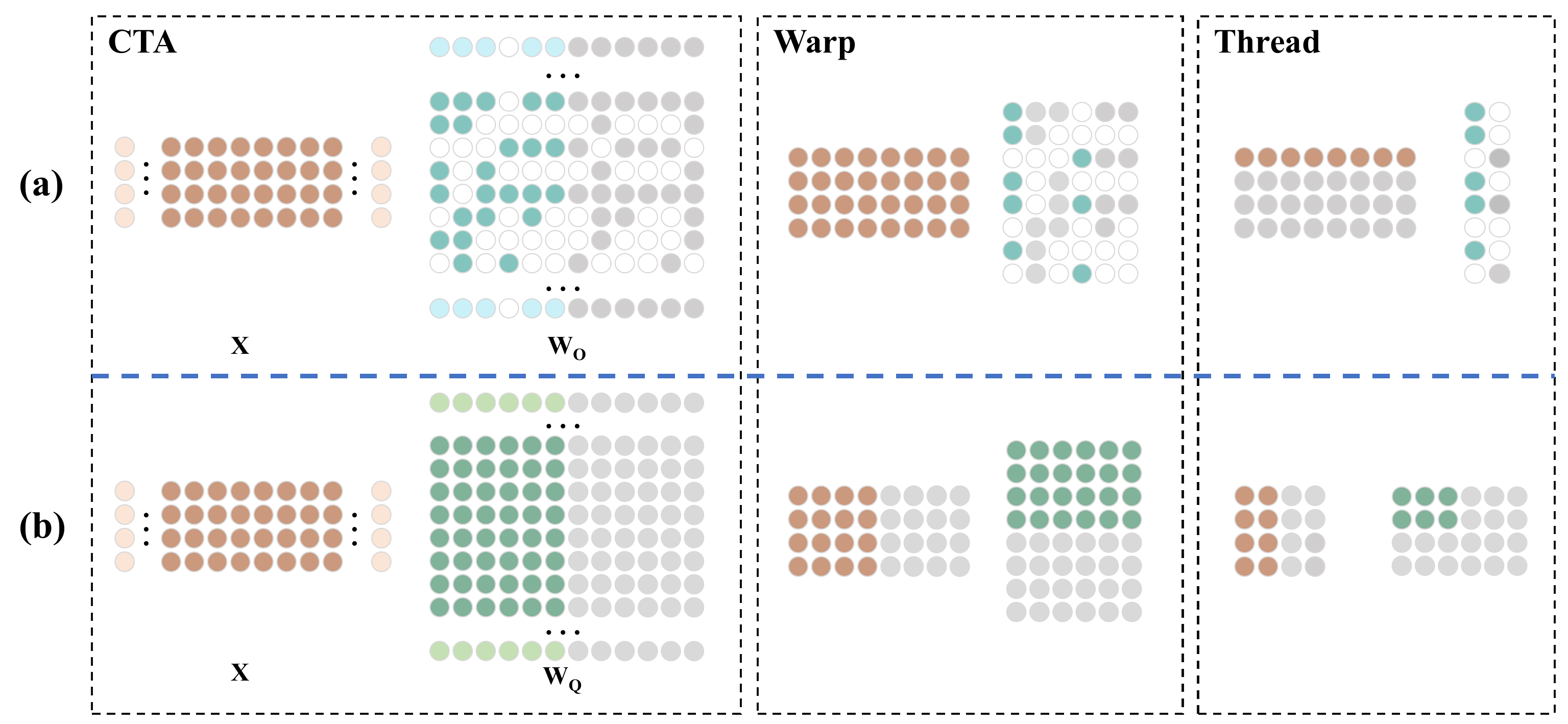}
    \caption{Example of the sparse–dense tiling. (a) Outlier tiling. (b) Quantized-weight tiling. Colored regions denote the active tiles. The example uses three warps per CTA (one dedicated to outliers) and eight threads per warp for clarity.}
    \label{fig:tiling}
\end{figure*}

\subsection{Sparse--Dense Tiling}
Content-sharing fusion requires tile correspondence: the dense and sparse paths must operate on compatible activation and output tiles. However, the conventional GEMM tile hierarchy cannot be directly reused for outlier SpMM because sparse entries are stored and accessed irregularly. To address this, {\toolName} constructs a shared sparse-dense tile hierarchy. The dense path preserves the CTA-, warp-, and thread-level tiling required by Tensor Core GEMM, while the sparse path is mapped to the corresponding regions of the activation and output tiles. Figure~\ref{fig:tiling} illustrates the resulting hierarchical partitioning and coordination between the two paths across CTA, warp, and thread levels.

\subsubsection{CTA Level}
At the CTA level, {\toolName} co-locates the dense and sparse computations that contribute to the same output tile. Specifically, the quantized weight matrix $W_Q \in \mathbb{R}^{K \times N}$ and the sparse outlier matrix $W_O \in \mathbb{R}^{K \times N}$ are partitioned into corresponding CTA-level tiles. 
For each CTA, the associated activation tile is loaded into shared memory once and reused by both the GEMM and SpMM paths. The partial results from the two paths are accumulated on chip before being written back to global memory, thereby avoiding redundant activation reloads and intermediate output traffic that would arise from separate-kernel execution.

This organization is particularly beneficial during decoding. 
Since the batch size $B$ is typically small, the activation matrix $X \in \mathbb{R}^{M \times K}$ is heavily reused across output columns, while weight reuse is comparatively limited. Accordingly, {\toolName} partitions the weight matrix across CTAs while sharing the activation tile within each CTA. Each CTA operates on
\[
X^{\text{CTA}} \in \mathbb{R}^{B \times K_{\text{CTA}}}, \quad
W_Q^{\text{CTA}} \in \mathbb{R}^{K_{\text{CTA}} \times N_{\text{CTA}}}, \quad
W_O^{\text{CTA}} \in \mathbb{R}^{K_{\text{CTA}} \times N_{\text{CTA}}}.
\]
However, naive partitioning along the N dimension can cause load imbalance when output tiles are not evenly distributed across SMs. To address this issue, {\toolName} adopts Stream-K, which distributes the aggregate MAC iteration space across persistent CTAs. Each CTA is assigned a contiguous segment of this space, potentially spanning multiple output tiles. This work-centric decomposition improves load balance while preserving activation reuse within the shared sparse–dense tile.

Each CTA-level tile is further decomposed into shared-memory tiles \(X^{SHM}\), \(W_Q^{SHM}\), and \(W_O^{SHM}\), whose sizes are selected to fit within the available shared-memory capacity. During execution, {\toolName} iteratively stages these subtiles from global memory to shared memory, allowing both dense and sparse computations to consume the same activation data through the on-chip memory hierarchy.

\subsubsection{Warp Level}
At the warp level, {\toolName} specializes warps according to the execution requirements of the dense and sparse computation paths. The dense path uses Tensor Core MMA instructions and thus requires regular, hardware-aligned fragments. In contrast, the sparse path runs on CUDA cores, as irregular outlier locations prevent efficient Tensor Core utilization.

For the dense path, each shared-memory tile $W_Q^{SHM}$ is partitioned into warp-level fragments. {\toolName} extends Sliced-K parallelism to the warp level, enabling multiple warps within the same CTA to compute independent K-fragments of the same output tile. The resulting partial sums are reduced within the CTA. Each dense warp processes
\[
X_Q^{\text{warp}} \in \mathbb{R}^{B \times K_{\text{warp}}}, \quad 
W_Q^{\text{warp}} \in \mathbb{R}^{K_{\text{warp}} \times N_{\text{warp}}}.
\]  

For the sparse path, the key challenge is balancing work while avoiding write conflicts. A uniform column-wise partition can assign highly uneven numbers of outliers to different warps, causing load imbalance. Conversely, a purely nonzero-balanced partition may cause multiple warps to update the same output columns, requiring atomic operations. {\toolName} avoids both issues by assigning each sparse warp a disjoint set of output columns while accounting for the local nonzero distribution. This design preserves conflict-free accumulation and improves sparse workload balance. Each sparse warp processes
\[
X_O^{\text{warp}} \in \mathbb{R}^{B \times K_{\text{CTA}}}, \quad 
W_O^{\text{warp}} \in \mathbb{R}^{K_{\text{CTA}} \times |C_{O,\text{warp}}|},
\]  
where $C_{O,\text{warp}}$ denotes the set of output columns assigned to the warp. Because sparse outlier processing is more susceptible to memory stalls than dense Tensor Core computation, {\toolName} allocates additional warps to the sparse path when needed, improving latency hiding through warp-level scheduling.

\subsubsection{Thread Level}
\label{sec:tiling_thread}
At the thread level, {\toolName} maps dense and sparse computations to distinct execution patterns while preserving a shared activation layout.

For GEMM, threads within a warp are mapped to the Tensor Core MMA tile layout (e.g., $16 \times 16 \times 8$), enabling efficient scheduling of quantized dense computations. In contrast, SpMM computations are executed on CUDA cores because their irregular access patterns limit Tensor Core utilization.

For dense computation, the activation matrix  $X^{SHM}$ is stored in shared memory in a \emph{row-major} layout with an 8-element swizzle (16~bytes, 4 banks) to reduce intra-subwarp bank conflicts. Under this layout, adjacent elements within the same column of \(X^{SHM}\) are spaced across 4 banks.
However, in SpMM, each thread reads only one element at a time due to the irregular sparsity pattern. Consequently,
bank conflicts occur when more than \(N_{\text{thread}} = \frac{\text{bank\_size}}{4}\) threads access the same column simultaneously. 
To mitigate this, each $W_O^{\text{warp}}$ is partitioned along the column dimension into \(N_{bucket} = \frac{\text{warp\_size}}{N_{thread}}\) buckets, each mapped to a group of \(N_{\text{thread}}\) threads.
Specifically, each thread handles  
\[
X_O^{\text{thread}} \in \mathbb{R}^{\tfrac{B}{N_{\text{thread}}} \times K_{\text{CTA}}}, \quad 
W_O^{\text{thread}} \in \mathbb{R}^{K_{\text{CTA}} \times |C_{O,\text{thread}}|},
\]  
where $\ C_{O,\text{thread}}$ denotes columns dynamically assigned to the bucket.

\subsection{Outlier Encoding}
\label{sec:encoding}
During workload partitioning, the outlier matrix is co-tiled with the quantized weight matrix, producing a coupled memory layout that is not well supported by conventional sparse formats such as CSR and COO. These formats are designed for standalone sparse kernels and do not explicitly preserve the tile-level locality required by fused sparse--dense GPU execution. In particular, the sparse representation in {\toolName} must satisfy four requirements: (1) alignment of outlier entries with GEMM tile boundaries, (2) vectorized access to sparse metadata and values, (3) direct compatibility with the shared-memory activation layout, and (4) balanced workload assignment under irregular outlier distributions. To meet these requirements, we introduce \texttt{\formatName}, a tile-local sparse format designed for efficient decoding within the sparse--dense tiling framework. We further apply offline reordering to improve load balance and reduce shared-memory bank conflicts.

\begin{figure}[h]
    \centering
    \includegraphics[width=0.95\linewidth]{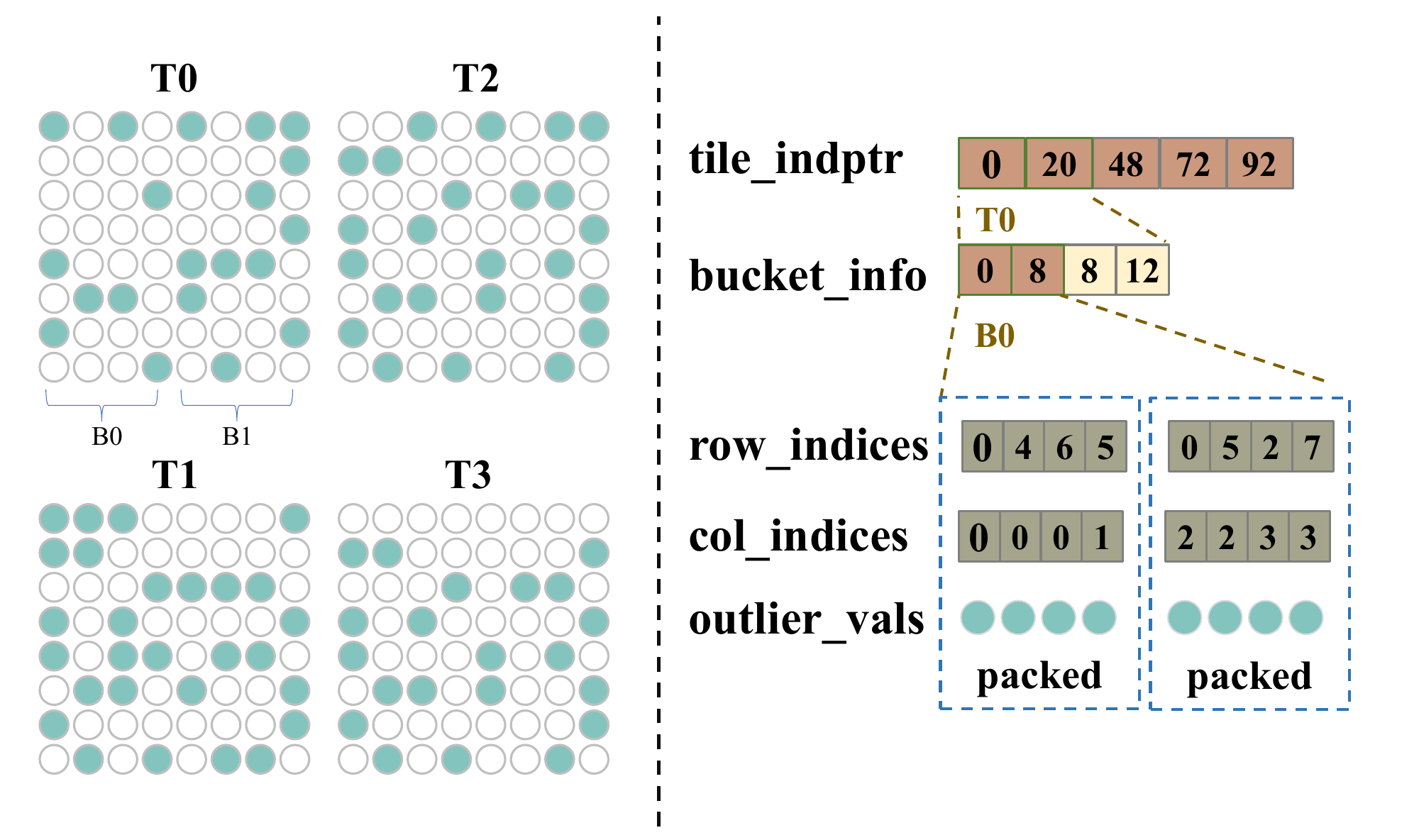}
    \caption{Example of {\formatName} with four tiles, each containing two buckets.}
    \label{fig:format}
\end{figure}

\subsubsection{\formatName{} Structure}
As illustrated in Figure~\ref{fig:format}, the \texttt{\formatName} format comprises three arrays: \texttt{tile\_indptr}, \texttt{bucket\_info}, and \texttt{packed\_outliers}. The \texttt{packed\_outliers} array stores the row indices, column indices, and values of the outlier elements. Overall, the format is designed to balance compression efficiency, GPU-friendly decodability, and compatibility with dense-tile reuse.

\paragraph{Tiles and Buckets.}
To align the sparse representation with the tiled execution of the quantized weight matrix, the outlier matrix is partitioned into shared-memory tiles that match the corresponding quantized weight tiles. The total number of \(W_O^{\mathrm{SHM}}\) tiles is
\[
N_{\mathrm{tile}} = \frac{K}{K_{\mathrm{SHM}}} \times \frac{N}{N_{\mathrm{SHM}}}.
\]
The \texttt{tile\_indptr} array stores the starting offset of each tile using \texttt{uint32\_t} indices. Within each tile, nonzero elements are further organized into column-oriented \emph{buckets} to improve memory locality. For each bucket, \texttt{bucket\_info} records both the starting offset and the number of nonzero elements. Since the number of elements per bucket is bounded, \texttt{uint16\_t} is sufficient to encode bucket sizes, thereby reducing storage overhead.

\paragraph{Packed Outliers.}
To support efficient decoding and vectorized memory access, nonzero elements within each bucket are packed into groups of four, with buckets containing fewer than four elements padded for alignment. During quantization, the unused padding slots can be repurposed to store original weight values, further improving model accuracy. Given the maximum shared-memory tile size of \(256 \times 256\), both \texttt{row\_index} and \texttt{col\_index} can be represented using \texttt{uint8\_t}, while outlier values are stored in \texttt{bfloat16}. Accordingly, every four outliers are packed into a \texttt{uint4}-based tuple consisting of row indices, column indices, and outlier values, enabling efficient parallel decoding during kernel execution.

\begin{table}[h]
    \centering
    \setlength{\tabcolsep}{4pt}
    \renewcommand{\arraystretch}{1.12}
    \caption{Data fields and storage costs in {\formatName}.}
    \label{tab:format_precision}
    \begin{tabular*}{\linewidth}{@{\extracolsep{\fill}} l l c c @{}}
        \toprule
        \textbf{Component} & \textbf{Field} & \textbf{Type} & \textbf{Count} \\
        \midrule
        \textit{tile\_indptr}
            & Tile offset
            & \texttt{UINT32}
            & \(N_{\mathrm{tile}} + 1\) \\
        \midrule
        \multirow{2}{*}{\textit{bucket\_info}}
            & Bucket offset
            & \texttt{UINT16}
            & \(N_B\) \\
            & Bucket nnz
            & \texttt{UINT16}
            & \(N_B\) \\
        \midrule
        \multirow{3}{*}{\textit{packed\_outliers}}
            & Row index
            & \texttt{UINT8}
            & \(\mathrm{nnz}\) \\
            & Column index
            & \texttt{UINT8}
            & \(\mathrm{nnz}\) \\
            & Outlier value
            & \texttt{BF16}/\texttt{FP16}
            & \(\mathrm{nnz}\) \\
        \bottomrule
    \end{tabular*}

    \vspace{2pt}
    \footnotesize
    \(N_B = N_{\mathrm{tile}} N_{\mathrm{warp}} N_{\mathrm{bucket}}\).
\end{table}

\paragraph{Storage Analysis.}
Table~\ref{tab:format_precision} summarizes the data types and sizes of all fields in {\formatName}. Accordingly, the total storage cost of {\formatName} is
\[
S_{\formatName}
= (N_{\mathrm{tile}}+1)\times 4
+ N_{\mathrm{B}} \times 4
+ \mathrm{nnz}\times 4
\;\text{bytes}.
\]
Compared with CSR, {\formatName} reduces both metadata overhead and decoding complexity by exploiting tile-level structure. Specifically, CSR requires a \((K+1)\)-entry row-pointer array and a 32-bit column index for each nonzero, yielding
\[
S_{\mathrm{CSR}}
= (K+1)\times 4
+ \mathrm{nnz}\times 12
\;\text{bytes}.
\]
By amortizing indexing metadata across tiles and buckets, {\formatName} shifts the overhead from per-element indexing to tile- and bucket-level descriptors, scaling with \(N_{\mathrm{tile}}\) and \(N_{\mathrm{tile}}\times N_{\mathrm{bucket}}\). This compact layout is well suited to sparse outlier distributions and GPU execution, as it improves memory locality and enables efficient parallel decoding.

\subsubsection{Sparse Reordering}
\label{sec:reordering}

To mitigate irregular outlier distributions, we adopt a two-stage sparse reordering strategy: column reordering improves workload balance across buckets, while element reordering reduces shared-memory bank conflicts.

\begin{figure}[h]
    \centering
    \includegraphics[width=0.9\linewidth]{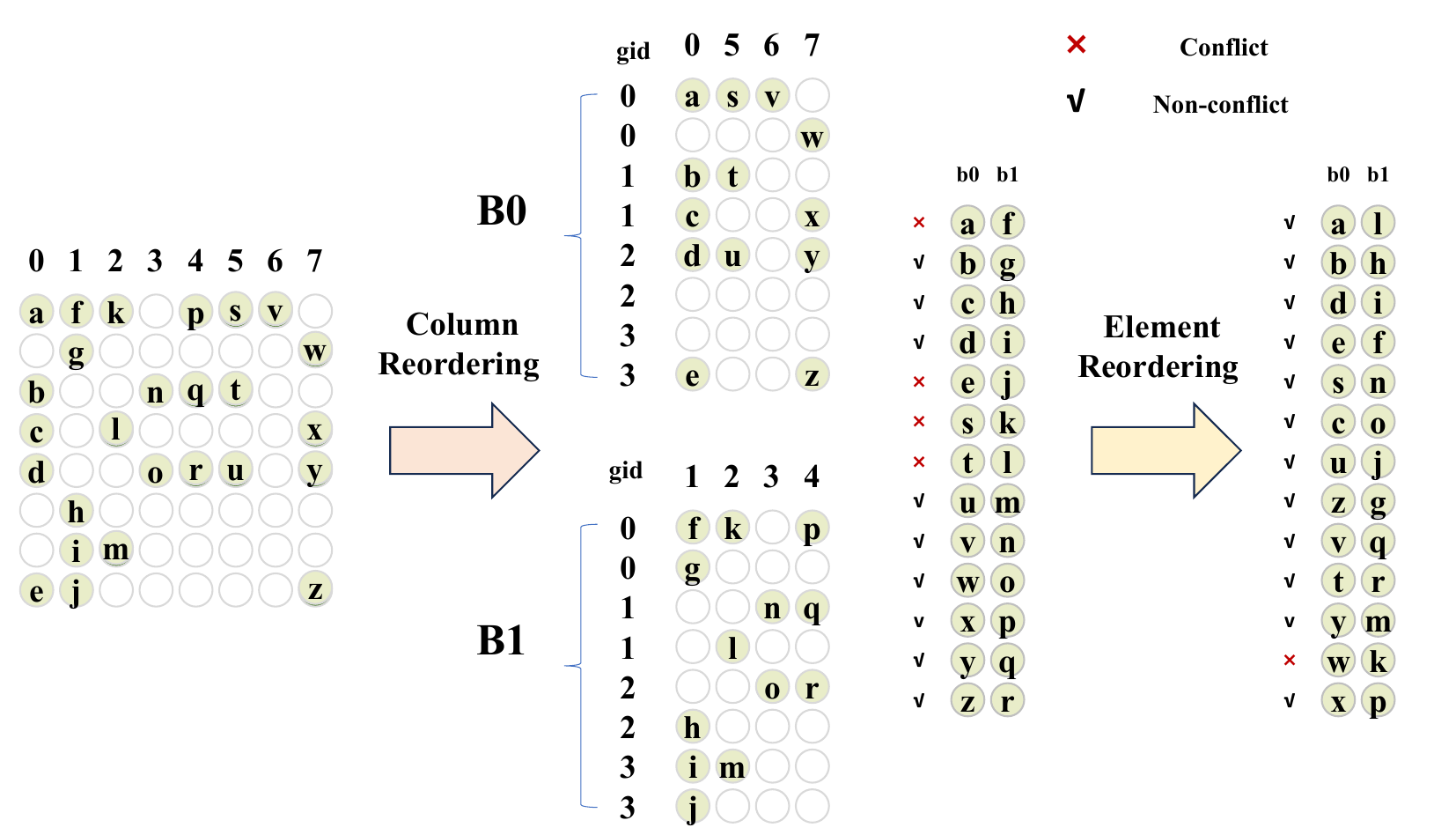}
    \caption{Example of column and element reordering with two buckets.}
    \label{fig:reordering}
\end{figure}

\paragraph{Column Reordering.}
Each shared tile \(W_O^{SHM}\) is partitioned into \(N_{warp} \times N_{bucket}\) buckets, where each bucket is processed by a dedicated sub-warp. Since columns may have highly skewed nonzero counts, assigning them in their original order can cause significant workload imbalance across sub-warps. To address this, we first sort the columns within each tile in descending order of nonzero count. We then use a greedy heap-based scheduling strategy: a min-heap maintains the cumulative nonzero count of each bucket, and each column is assigned to the currently least-loaded bucket. This procedure balances the nonzero distribution across buckets and reduces inter-bucket workload imbalance. As shown in Figure~\ref{fig:reordering}, the eight columns in the tile are distributed across two buckets with approximately balanced nonzero counts.

\paragraph{Element Reordering.}
\label{sec:element_reordering}
Under our thread-to-data mapping, adjacent threads within a sub-warp access shared memory with a stride of four banks, as described in Section~\ref{sec:tiling_thread}. Consequently, accesses from different sub-warps tend to follow a four-bank-group pattern, where target banks are grouped into four consecutive banks, such as banks 0--3 and 4--7. Bank conflicts arise when multiple sub-warps simultaneously access elements mapped to the same bank group. Since each sub-warp processes one bucket, we reduce such conflicts by reordering the nonzero elements within each bucket according to their target bank groups. Specifically, each element is assigned to one of four groups as
\[
    \text{gid} = \left\lfloor \frac{k}{2} \right\rfloor \% 4,
\]
where \(k\) denotes the row index. Elements in each bucket are then selected following the bucket-dependent circular order
\[
    \left([0, 1, 2, 3] + \text{bucket\_id}\right) ~\%~ 4.
\]
This ordering staggers accesses from different sub-warps across bank groups, thereby reducing inter-sub-warp bank conflicts. As shown in Figure~\ref{fig:reordering}, this strategy reduces the number of conflicts from 4 to 1.

\begin{listing}[b]
\begin{minted}[frame=single, framesep=2mm, fontsize=\footnotesize, escapeinside=||, linenos, keywordcase=upper]{cuda}
/* Input: 
    Packed outliers: outliers
    Swizzles: swizzles
*/
// Thread Configuration
int bucket_id = threadIdx.x / 8;
int row = threadIdx.x % 8;
scalar_t* sh_x_row = sh_x + row * X_SHARE_STRIDE;
uint16_t bucket_start = outlier_ptr[bucket_id*2];
uint16_t bucket_nnz = outlier_ptr[bucket_id*2+1];
outlier_ptr += bucket_start;

// Double buffer
uint4 outliers[2];
int buffer_id = 0;
outliers[buffer_id] = outlier_ptr[0];

for (int i = 0; i < bucket_nnz; i++) {
    // Unpack
    if (i < bucket_nnz - 1)   
        outliers[buffer_id ^ 1] = outlier_ptr[i+1];
    uint4 outlier = outliers[buffer_id];
    buffer_id ^= 1;
    uint32_t ks, col;
    ks = reinterpret_cast<uint32_t*>(&outlier)[0];
    col = reinterpret_cast<uint32_t*>(&outlier)[1];
    scalar_t* vals = reinterpret_cast<scalar_t*>(&outlier);
    
    // Apply swizzle once
    ks ^= swizzles;
    
    // Computation
    #pragma unroll
    scalar_t res = 0.0;
    for (int i = 0; i < 4; i++) {
        uint8_t k = reinterpret_cast<uint8_t>(&ks)[i];
        scalar_t X_val = sh_x_row[k];
        scalar_t O_val = vals[i];
        res += X_val * O_val;
    }
    sh_c[row, col] = res;
    
}    

\end{minted}
\caption{Sparse Outlier Decoding Procedure.
}
\label{alg:decoding}
\end{listing}

\section{Implementation}

This section presents two key components of the {\toolName} implementation. We first describe a pipelined execution strategy that overlaps computation with data movement to hide memory-access latency. We then introduce an optimized outlier decoding mechanism that improves sparse--dense computation efficiency while preserving overall performance.

\subsection{Efficient Pipelined Execution}

\begin{figure}[ht]
    \centering
    \includegraphics[width=0.9\linewidth]{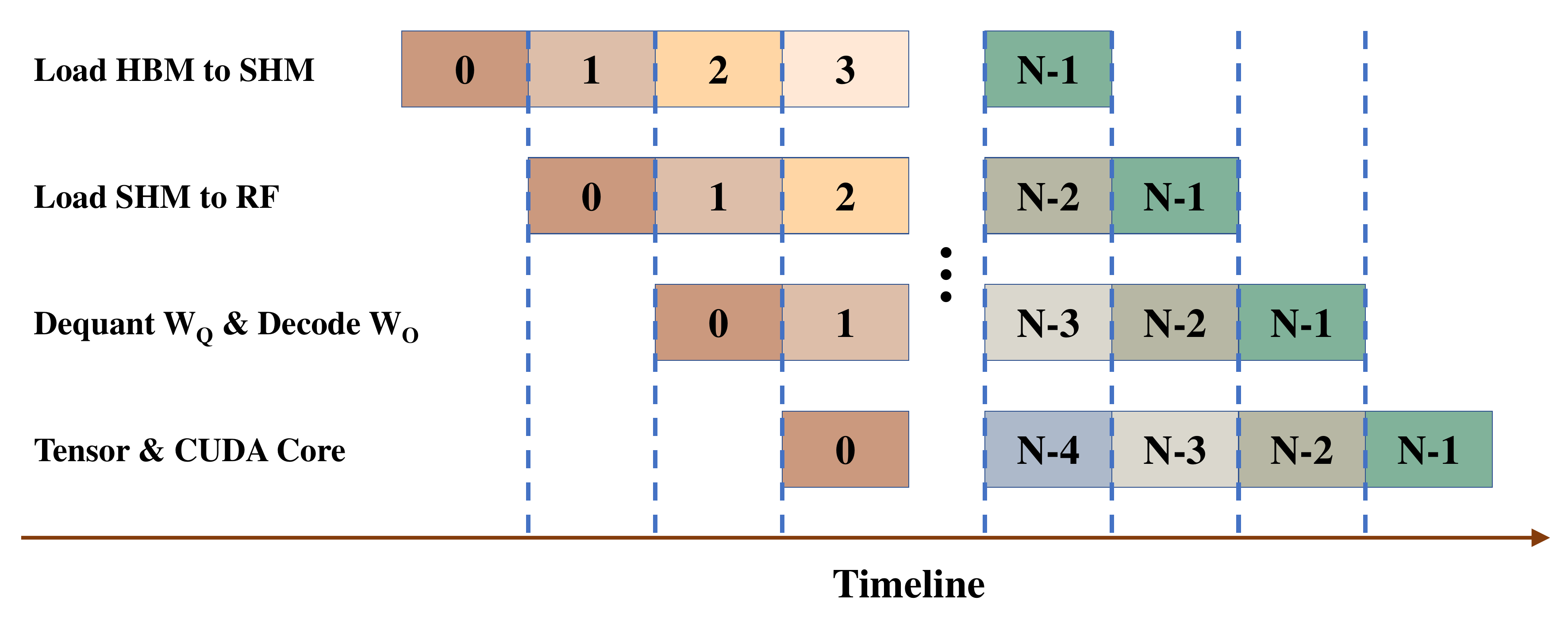}
    \caption{Illustration of the pipelined execution strategy. Blocks with the same color denote identical shared tiles.}
    \label{fig:pipeline}
\end{figure}

As shown in Figure~\ref{fig:pipeline}, {\toolName} adopts a multi-stage pipelined execution model with double buffering to overlap data movement and computation, thereby reducing global-memory latency and improving hardware utilization. The operator consists of four stages: (1) loading data from global memory, (2) staging data in registers, (3) performing dequantization and outlier decoding, and (4) executing computation on Tensor Cores and CUDA Cores.

In the first stage, the quantized weight matrix $W_Q$, sparse outlier matrix $W_O$, and activation matrix $X$ are transferred from global memory to shared memory using asynchronous copy operations. In the second stage, each thread loads the required quantized weights, activations, and compressed outlier values from shared memory into registers. Because outlier processing has relatively low arithmetic intensity and is therefore more sensitive to memory-access latency, we allocate additional warps to this task, enabling latency hiding through warp-level context switching.
During computation, each thread dequantizes the dense quantized weights, multiplies them with the corresponding activations, and stores intermediate results in registers. In parallel, each thread decodes outlier elements, fetches the corresponding activations from shared memory, and accumulates the resulting outlier contributions in shared-memory buffers. After all tiles are processed, these contributions are reduced into registers. When required by the Stream-K strategy, a subsequent global reduction is performed to produce the final outputs. The results are then written back to global memory.

\begin{figure*}[b]
    \centering
    \begin{subfigure}[t]{0.29\textwidth}
        \centering
        \includegraphics[width=\linewidth]{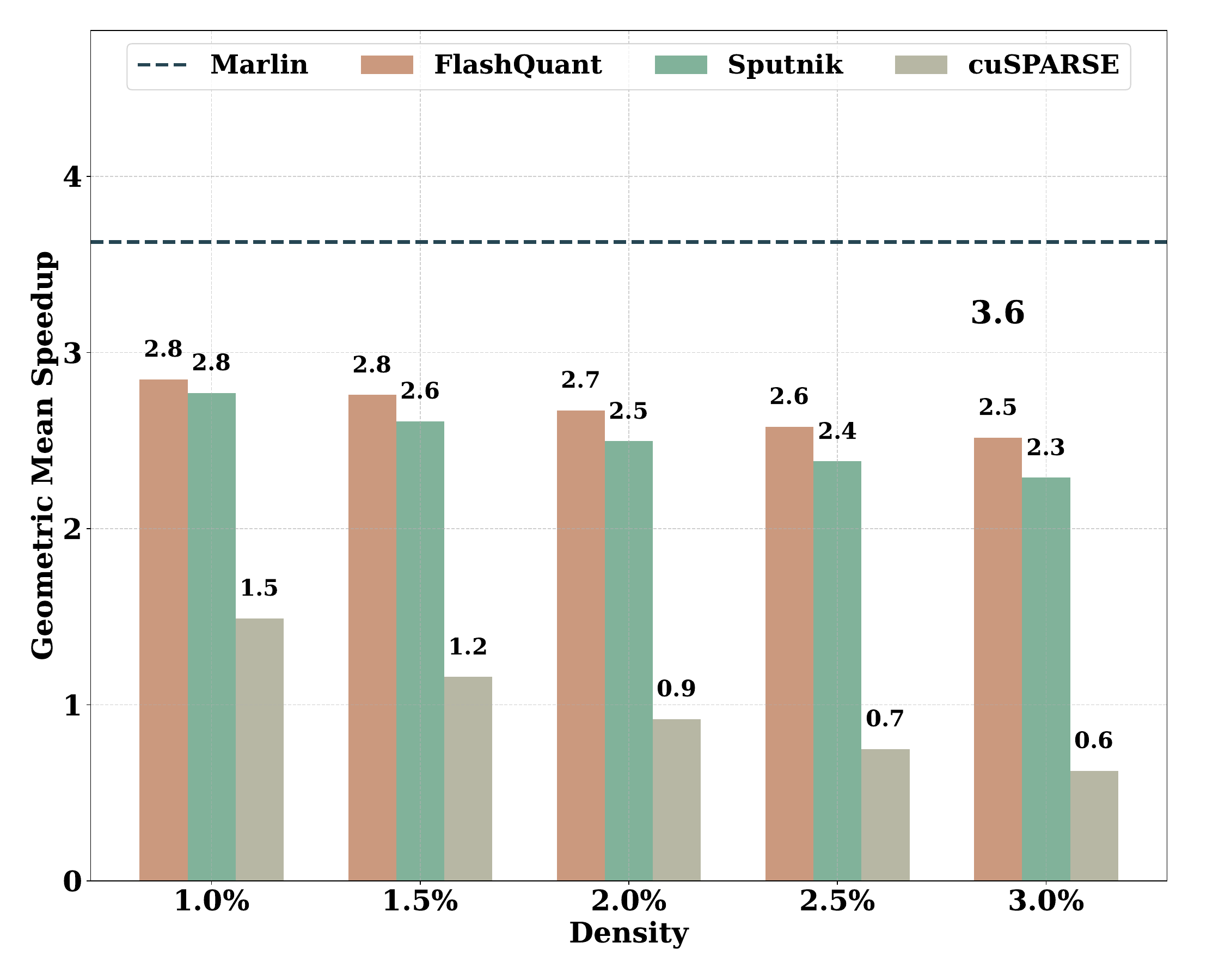}
        \caption{Batch Size = 8, RTX 3090}
    \end{subfigure}
    \begin{subfigure}[t]{0.29\textwidth}
        \centering
        \includegraphics[width=\linewidth]{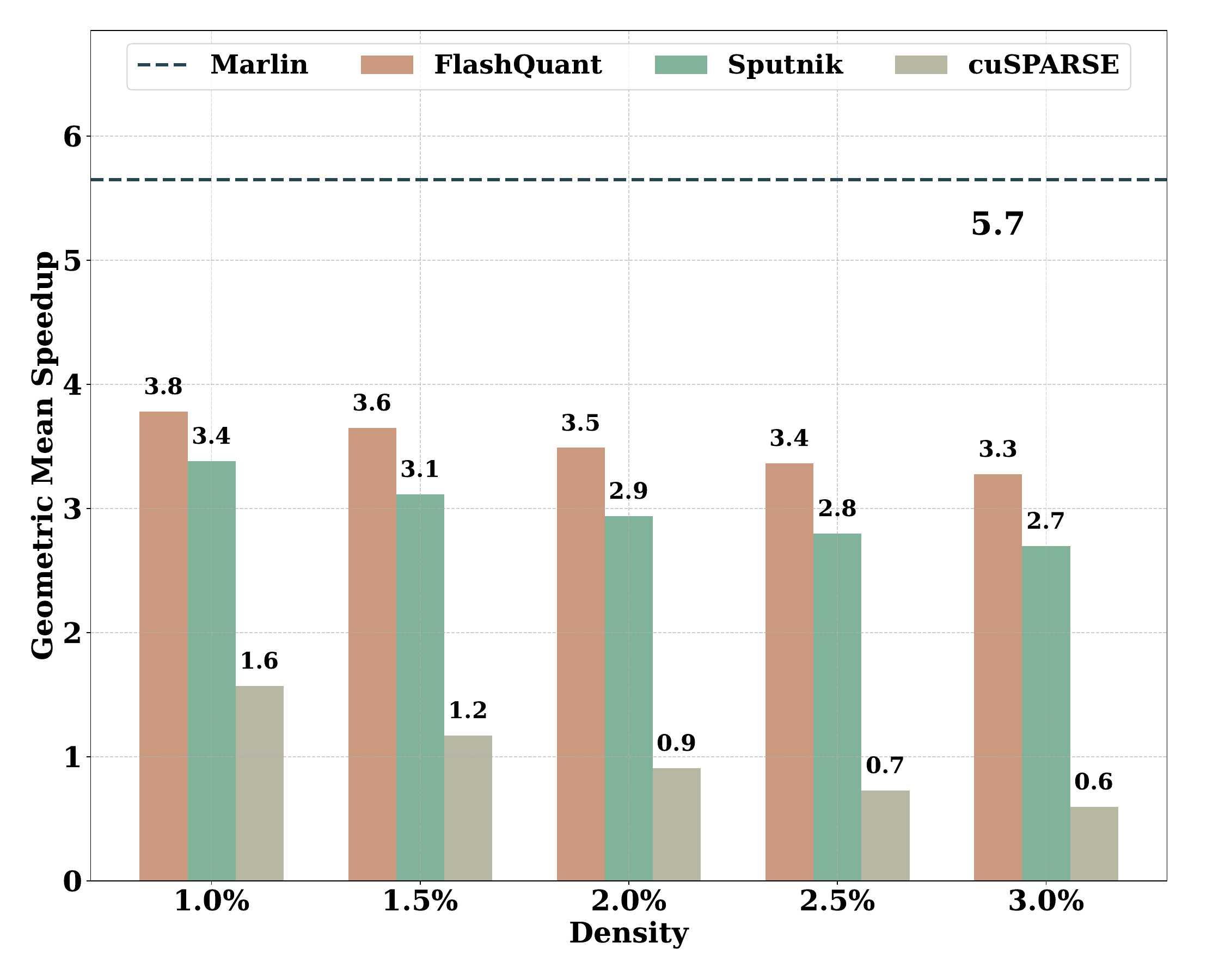}
        \caption{Batch Size = 8, RTX 4090}
    \end{subfigure}
    \begin{subfigure}[t]{0.29\textwidth}
        \centering
        \includegraphics[width=\linewidth]{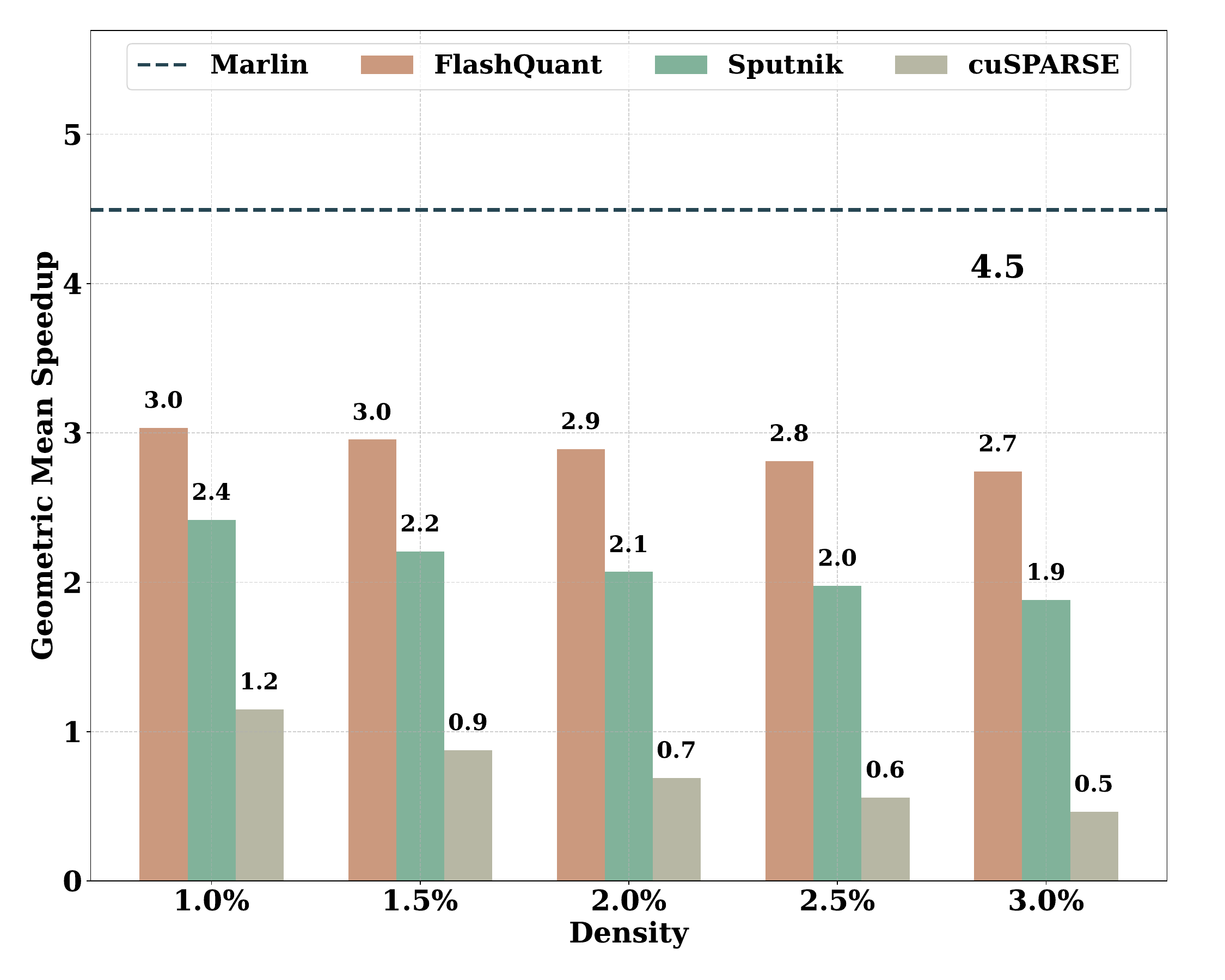}
        \caption{Batch Size = 8, RTX 5090}
    \end{subfigure}

    \vspace{0.5em}

    \begin{subfigure}[t]{0.29\textwidth}
        \centering
        \includegraphics[width=\linewidth]{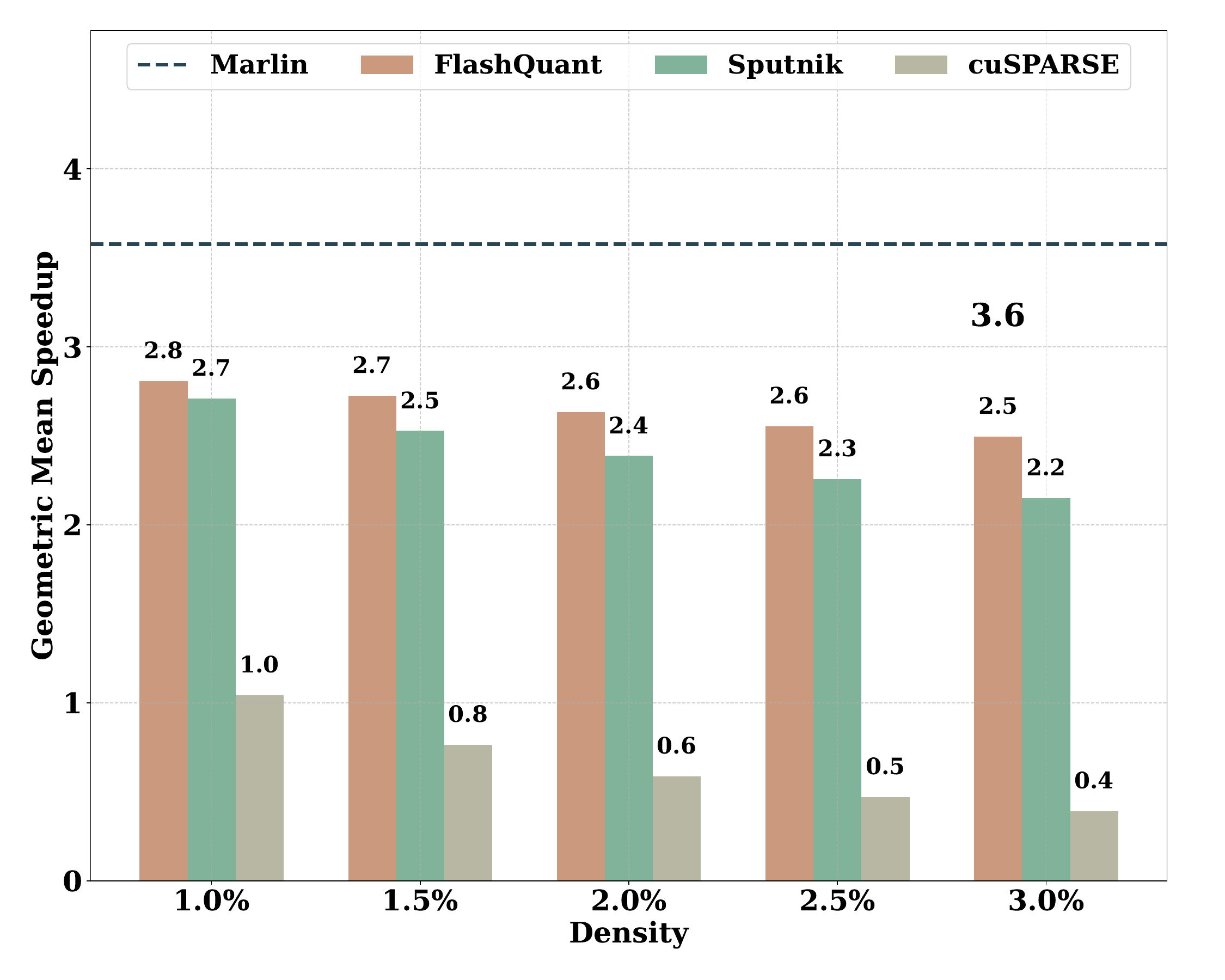}
        \caption{Batch Size = 16, RTX 3090}
    \end{subfigure}
    \begin{subfigure}[t]{0.29\textwidth}
        \centering
        \includegraphics[width=\linewidth]{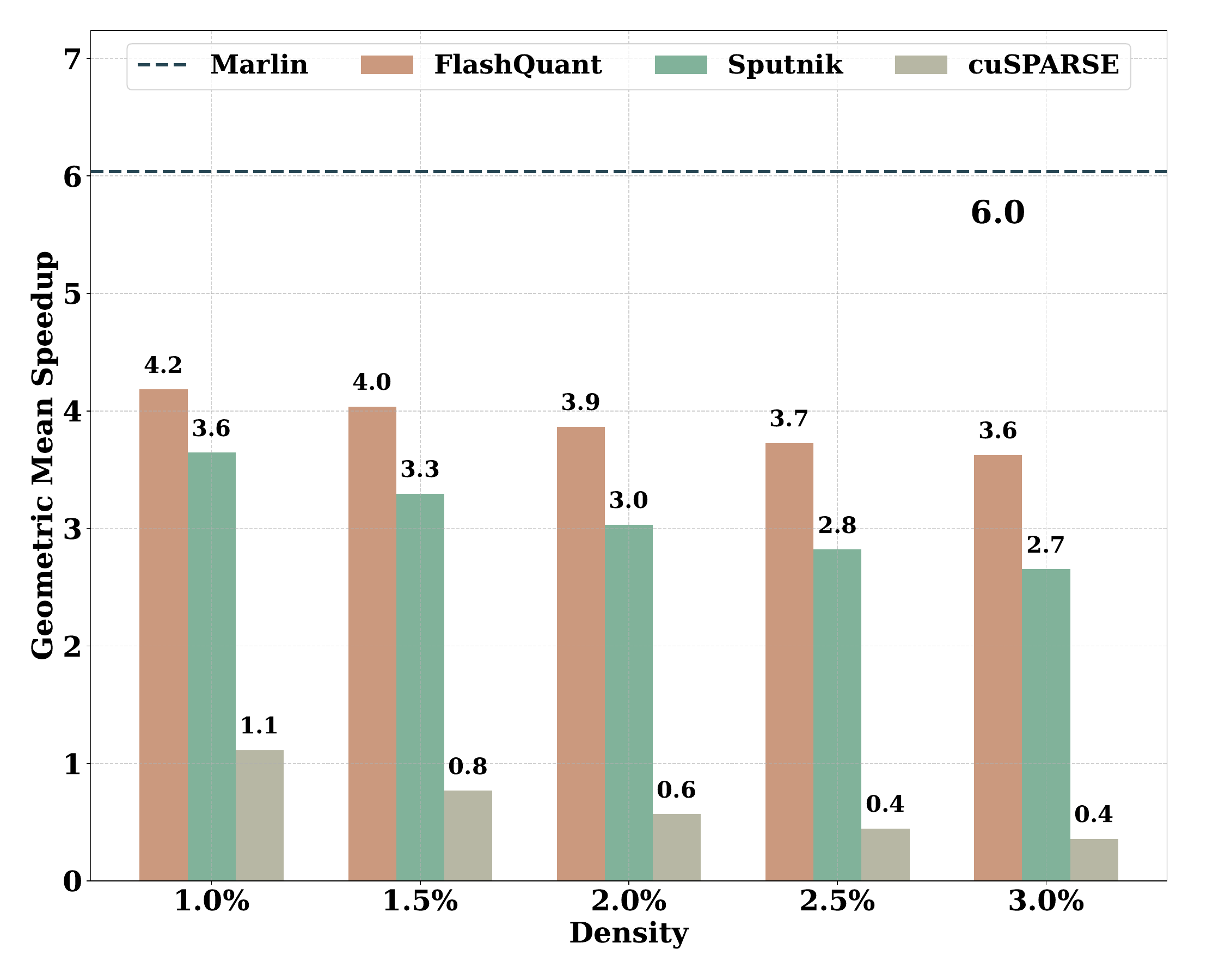}
        \caption{Batch Size = 16, RTX 4090}
    \end{subfigure}
    \begin{subfigure}[t]{0.29\textwidth}
        \centering
        \includegraphics[width=\linewidth]{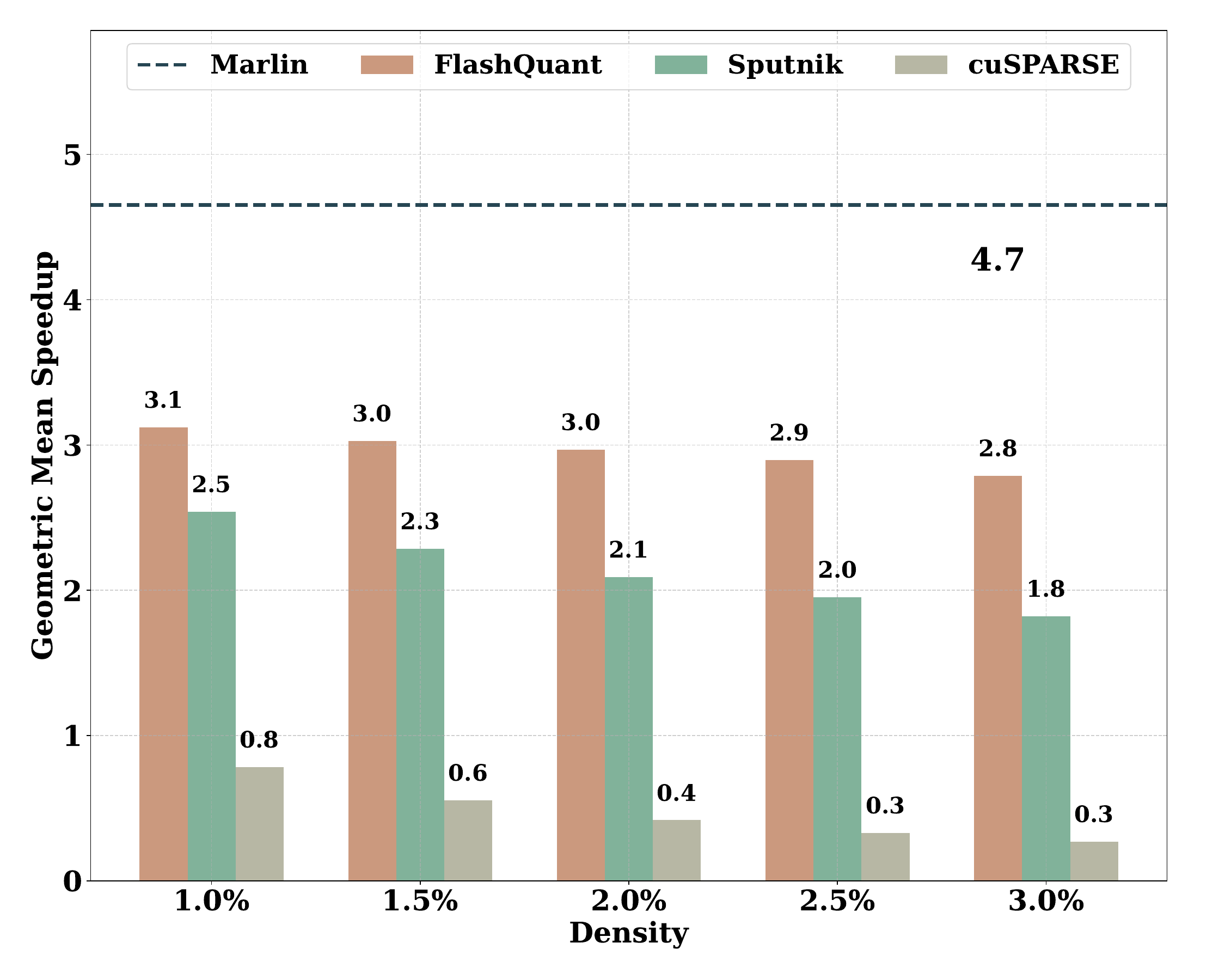}
        \caption{Batch Size = 16, RTX 5090}
    \end{subfigure}

    \caption{Speedup of various outlier-aware quantization kernel implementations compared with cuBLAS (BF16).}
    \label{fig:kerperf}
\end{figure*}

\subsection{High-Performance Outlier Decoding}
\label{sec:decoding}

Efficient SpMM execution requires decoding the outlier matrix into a representation compatible with the kernel's shared-memory layout. Listing~\ref{alg:decoding} illustrates the GPU-optimized decoding procedure used to reconstruct outlier entries prior to computation.

Each thread computes its \texttt{bucket\_id} and the corresponding address in the shared-memory activation matrix (lines~6--8). For each bucket, the kernel loads the starting offset \texttt{bucket\_start} and the number of nonzero outliers \texttt{bucket\_nnz}, then processes these entries sequentially (lines~9--11). To hide global-memory latency, the implementation uses double buffering, prefetching the next outlier entry while decoding the current one (lines~14--15 and 20--23).

Each packed outlier is decoded into three components: the $k$ indices \texttt{ks}, a column index \texttt{col}, and four half-precision values \texttt{vals} (lines~24--27). To improve vectorization and register reuse, the number of outliers per column is padded to a multiple of four, allowing four values from the same column to share a single encoded column index. The k indices are then transformed by a swizzle operation to match the shared-memory layout of the activation matrix~$X$ (line~30). The mask variable \texttt{swizzles}, consisting of four \texttt{uint8\_t} entries, is precomputed once and reused across the four elements, reducing per-thread overhead.

After decoding, each thread loads the corresponding activation values from shared memory using the swizzled indices and executes four unrolled fused multiply--add (FMA) operations (lines~33--40). The resulting partial sums are written back to shared memory according to the \texttt{row} and \texttt{col} indices (line~41).

\begin{table}[h]
\caption{Hardware Configurations.}
\label{tab:hardware}

\begin{tabular}{crrr}
   \toprule
   GPU               & \thead{CUDA Cores \\ (FP16/BF16)}                 & \thead{Tensor Cores \\ (FP16/BF16)} & Bandwidth\\ \midrule
RTX 3090 & 35.6 TFLOPS & 71.0 TFLOPS & 936 GB/s \\
RTX 4090 & 82.6 TFLOPS & 165.2 TFLOPS & 1008 GB/s \\
RTX 5090 & 104.8 TFLOPS & 209.5 TFLOPS & 1792 GB/s \\    \bottomrule
\end{tabular}

\end{table}

\section{Evaluation}

\subsection{Setup}
\textbf{Hardware.}
We evaluate {\toolName} on three representative consumer GPUs, the NVIDIA GeForce RTX 3090, RTX 4090, and RTX 5090, which span three architectural generations.
This selection enables a consistent assessment of the scalability and portability of our optimizations across evolving GPU microarchitectures. Table~\ref{tab:hardware} summarizes the relevant hardware characteristics, including peak FP16/BF16 throughput for CUDA Cores and Tensor Cores, as well as sustained memory bandwidth.

\noindent
\textbf{Datasets.}
We extract representative quantization operator shapes from several LLM families, including LLaMA~\cite{grattafiori2024llama}, Qwen~\cite{team2024qwen2}, DeepSeek \cite{liu2024deepseek}, and OPT~\cite{zhang2022opt}. 
For each operator shape, we generate five outlier ratios: \{1.0\%, 1.5\%, 2.0\%, 2.5\%, 3.0\%\}. 
This yields $340$ \texttt{INT4+BF16} outlier-aware operator configurations. 
The resulting problem sizes range from $(4096, 4096)$ to $(75264, 14848)$.

\noindent
\textbf{Baselines.}
We compare {\toolName} against strong unfused baselines built from state-of-the-art dense and sparse kernels. To the best of our knowledge, prior systems do not provide a fused operator for the mixed dense–sparse computation pattern required by outlier-aware quantized inference. Therefore, in existing systems, the practical deployment strategy is to execute the dense quantized path and the sparse outlier path separately. Following this design, we combine \textbf{Marlin} for dense quantized GEMM with either \textbf{Sputnik} or \textbf{cuSPARSE} for sparse outlier processing, and we account for the full cost of the unfused pipeline. We also report \textbf{cuBLAS} BF16 as a high-precision reference.
Methods such as MixQ~\cite{chen_mixq_nodate}, FLUTE~\cite{guo_fast_2024}, and framework-native quantization paths differ in outlier handling or execution formulation and are therefore discussed in related work rather than used as the main system baselines.

\subsection{Kernel Benchmark}

Figure~\ref{fig:kerperf} compares {\toolName} with the baselines at batch sizes of 8 and 16. We include Marlin as an upper-bound reference for the outlier-aware kernel, corresponding to an idealized setting in which sparse outlier handling is ignored. Relative to cuBLAS BF16, {\toolName} achieves speedups of $2.74\times$--$4.18\times$, primarily due to the benefits of quantized computation. Compared with the strongest unfused baseline, Sputnik+Marlin, {\toolName} achieves speedups of $1.03\times$--$1.16\times$ on the RTX~3090, $1.12\times$--$1.37\times$ on the RTX~4090, and $1.23\times$--$1.53\times$ on the RTX~5090.

\begin{figure}[h]
    \centering    
    \includegraphics[width=0.8\linewidth]{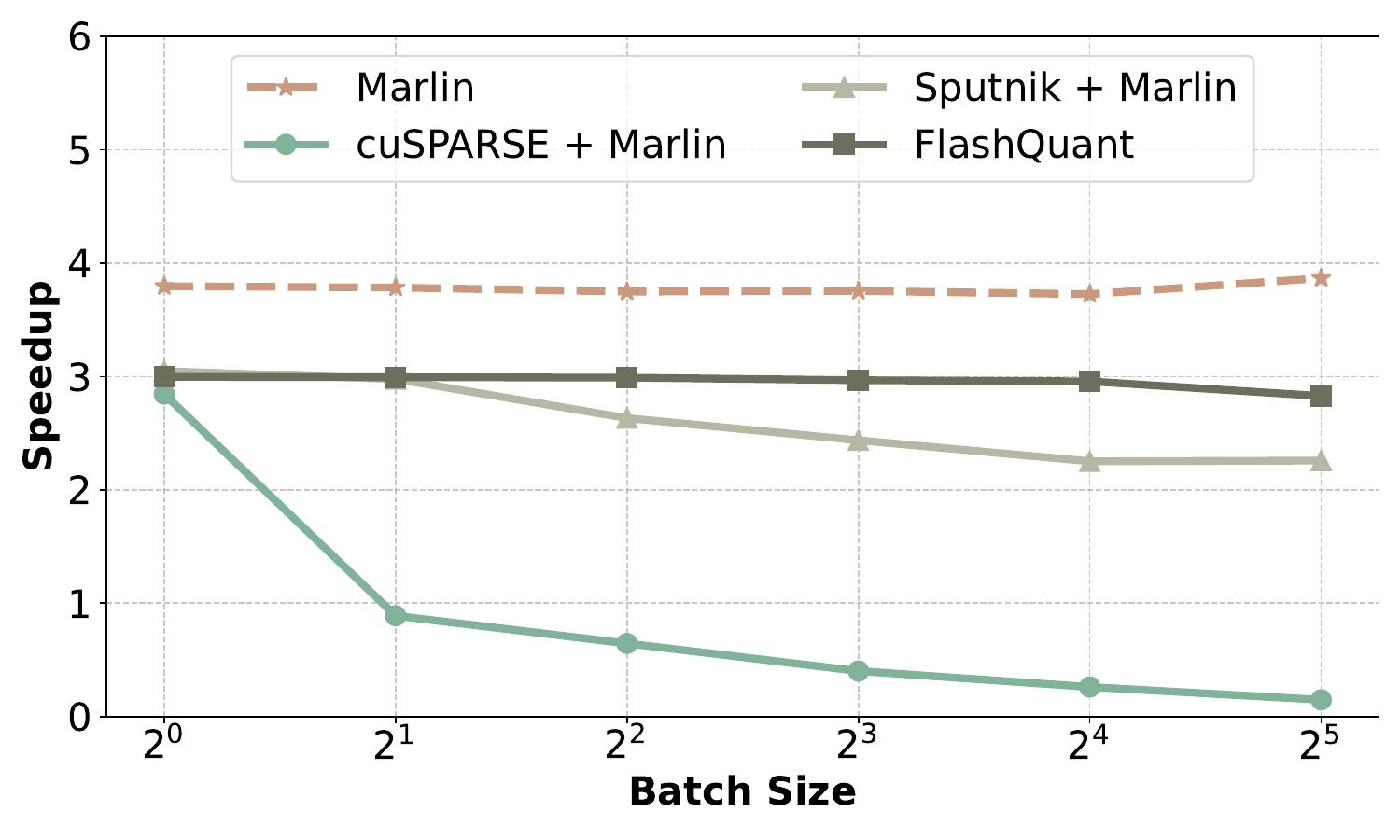}
    \caption{Speedup over cuBLAS BF16 on RTX~4090 for kernel $(14848, 75264)$ with $3\%$ outlier density.}
    \label{fig:batch}
\end{figure}

Beyond these overall speedups, Figure~\ref{fig:kerperf} reveals two additional trends. First, as outlier density increases, {\toolName} exhibits a smaller latency increase than Sputnik, demonstrating the benefit of fused sparse--dense execution in denser regimes. Second, the advantage of {\toolName} becomes more pronounced on newer GPUs with higher floating-point throughput. Although fusion introduces additional computation, this overhead becomes relatively smaller on newer architectures, while the reduction in memory traffic continues to provide performance benefits.

Figure~\ref{fig:batch} evaluates performance across batch sizes. {\toolName} consistently outperforms the baselines at a batch size of 32, indicating good scalability to larger workload granularities. For batch sizes below 8, however, {\toolName} pads inputs to a minimum size of 8, introducing redundant computation and reducing parallel efficiency. Consequently, its advantage diminishes for very small batches, such as 1 and 2.

\subsection{Cost Analysis}

\begin{figure}[h]
    \centering
    \includegraphics[width=0.85\linewidth]{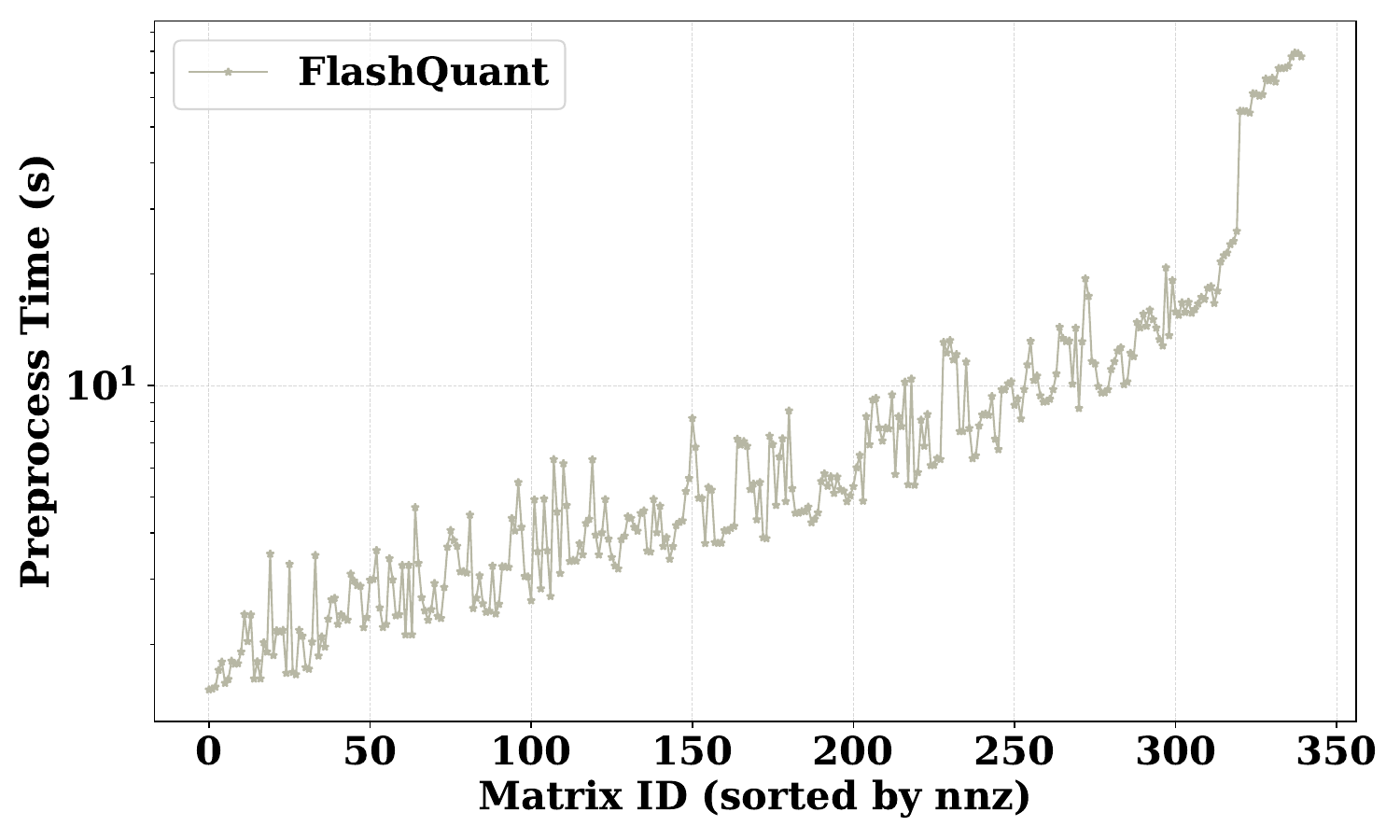}
    \caption{Preprocessing time of {\toolName}.}
    \label{fig:p_cost}
\end{figure}

\begin{figure}[h]
    \centering
    \includegraphics[width=0.9\linewidth]{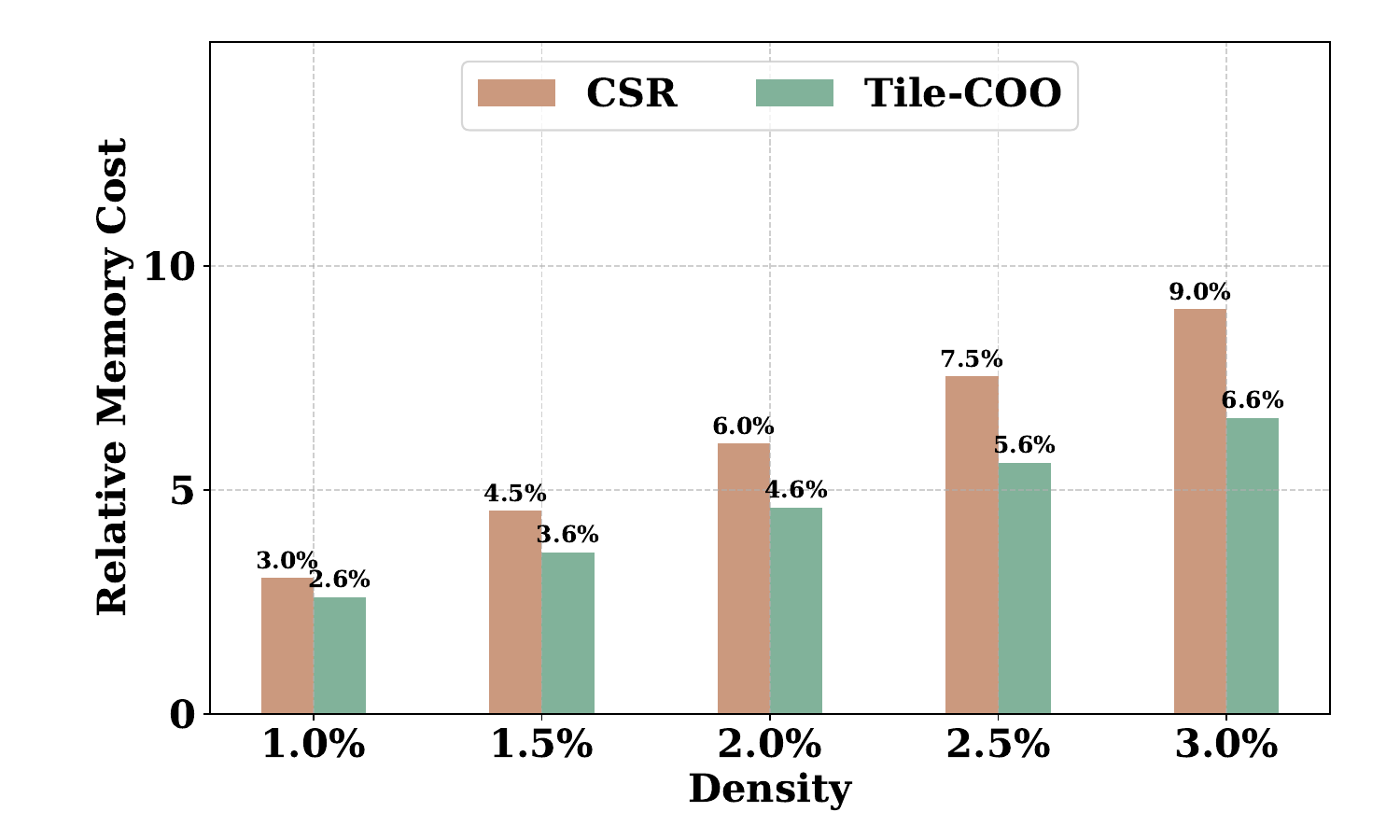}
    \caption{Memory overhead of outlier matrices.}
    \label{fig:m_cost}
\end{figure}

We report the preprocessing time and memory overhead of the proposed method. Preprocessing comprises reordering and packing, and Figure~\ref{fig:p_cost} shows that the total preprocessing time remains below 100~s for all evaluated matrices. Because this cost is incurred only once at deployment, it is amortized across subsequent executions and does not affect steady-state performance. We further report the storage overhead normalized to the dense format across varying matrix densities in Figure~\ref{fig:m_cost}. The overhead is dominated by metadata, primarily indices, whose storage cost often exceeds that of the outlier values. By employing low-precision index encoding with 8-bit row and column indices, the {\formatName} format reduces metadata size and consequently requires less storage than CSR.

\subsection{Ablation Study}
\begin{figure}[t]
    \centering
    \includegraphics[width=0.9\linewidth]{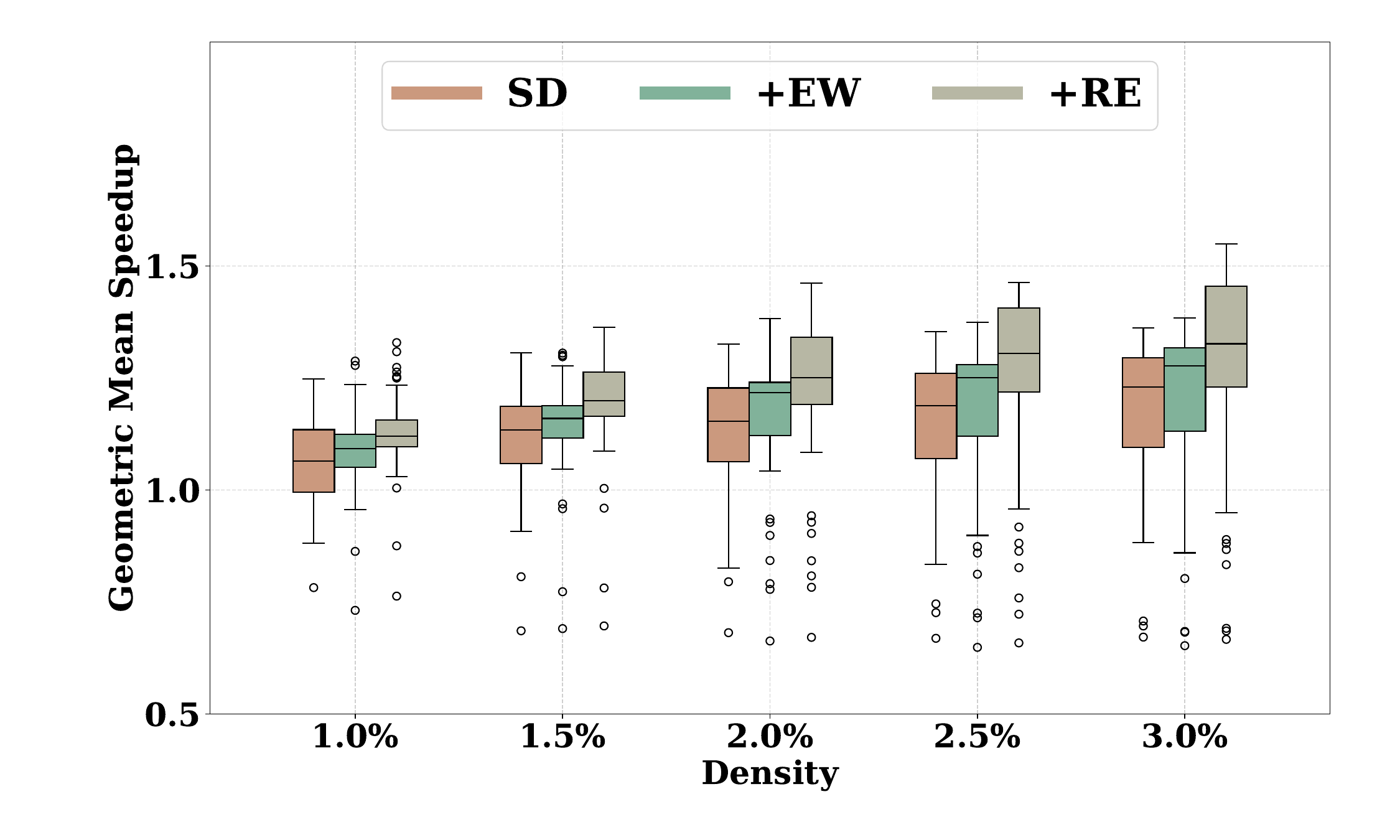}
    \caption{Effect of various optimizations.}
    \label{fig:ablation}
\end{figure}

\begin{figure}[h]
    \centering
    \includegraphics[width=0.8\linewidth]{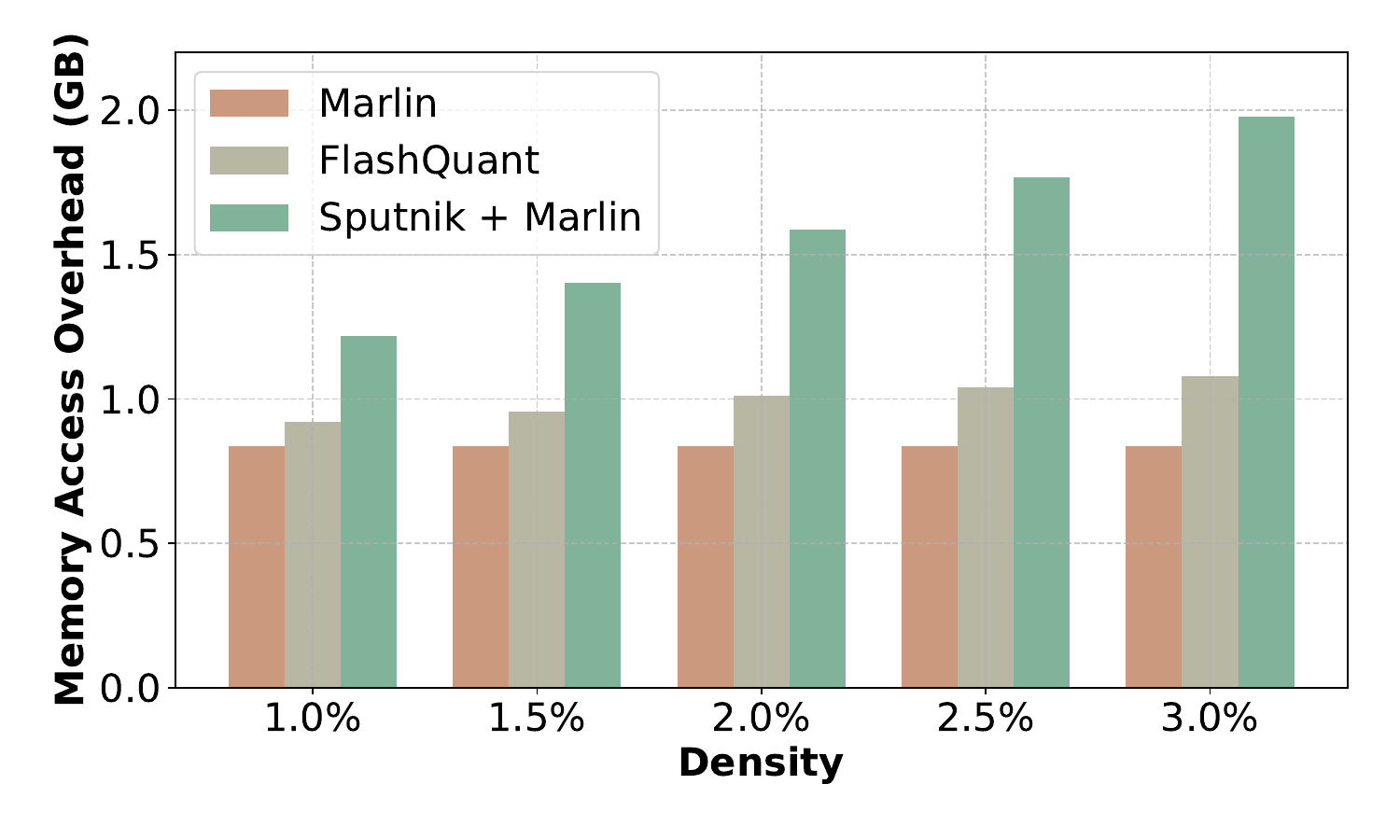}
    \caption{ Memory access overhead for a $(16,14848)\times(14848,75264)$ kernel on an RTX~4090 GPU.}
    \label{fig:batch_memory}
\end{figure}

We perform an ablation study by incrementally enabling three techniques: sparse--dense tiling (\textbf{SD}), extra warp allocation (\textbf{EW}), and reordering (\textbf{RE}). Figure~\ref{fig:ablation} reports the geometric mean speedups over Sputnik+Marlin on the RTX~4090. With only SD enabled, {\toolName} achieves a 6\%--16\% speedup over Sputnik+Marlin. This improvement stems from co-tiling sparse outlier processing with dense computation, which enhances on-chip reuse of activations and partial outputs while eliminating the additional global memory accesses incurred by the sparse SpMM kernel. Figure~\ref{fig:batch_memory} further shows that {\toolName} reduces memory traffic by up to 45.4\% compared with Sputnik. Since outlier SpMM has low arithmetic intensity, we allocate dedicated warps to improve latency hiding, resulting in an additional 2\%--4\% speedup. Finally, to mitigate load imbalance caused by irregular sparsity, we reorder columns to distribute high-density columns more evenly across warps and reorder elements to reduce shared-memory bank conflicts. Enabling RE yields a further 4\%--9\% speedup.

\subsection{End-to-End}

\begin{table}[h]
\small
\caption{Perplexity on the WikiText2 dataset.}
\label{tab:perplexity}
\resizebox{\linewidth}{!}{%

\begin{tabular}{cccccc}
\toprule
\multirow{2}{*}{Method} & \multirow{2}{*}{Threshold} & \multicolumn{2}{c}{LLaMA3-8B} & \multicolumn{2}{c}{Qwen2.5-14B} \\
                        &                            & Memory (GiB)      & PPL       & Memory (GiB)       & PPL        \\ \midrule
Baseline                       & -                          & 14.96             & 5.538     & 27.51              & 4.741      \\
GPTQ                           & -                          & 5.34              & 6.896     & 9.30               & 5.191      \\
\multirow{4}{*}{SpQR} & 5\%                        & 6.06              & \textbf{5.925}     & 10.73              & \textbf{5.092}      \\
                               & 10\%                       & 5.83              & 5.936     & 10.26              & 5.120      \\
                               & 15\%                       & 5.75              & 5.966     & 10.10              & 5.145      \\
                               & 20\%                       & 5.70              & 5.979     & 10.01              & 5.152     
\\ \bottomrule
\end{tabular}
   
}
\end{table}

We evaluate {\toolName} in an end-to-end setting on LLaMA3-8B~\cite{grattafiori2024llama} and Qwen2.5-14B~\cite{team2024qwen2}. We compare the original models, GPTQ-quantized models, and SpQR-based outlier-aware quantized models~\cite{dettmers_spqr_2023} across four quantization thresholds, where higher thresholds retain fewer outliers. Table~\ref{tab:perplexity} reports WikiText2 perplexity and model size. Compared with GPTQ, outlier-aware quantization consistently reduces perplexity with only modest memory overhead.

We further evaluate four vLLM configurations~\cite{kwon2023efficient}. \textbf{BF16} denotes the baseline using the original vLLM implementation with BF16 precision, while \textbf{Marlin} uses the Marlin backend with GPTQ-quantized models. \textbf{cuSPARSE} and \textbf{{\toolName}} are alternative implementations for outlier-aware quantized models, with the quantization threshold fixed at 5\%. Table~\ref{tab:inference} reports decoding throughput. {\toolName} achieves a $2.01 \times$ end-to-end speedup over the BF16 baseline at batch size 8 and a $1.98 \times$ speedup at batch size 16. Moreover, Qwen2.5 does not fit in GPU memory under the BF16 baseline, whereas quantization sufficiently reduces the memory footprint to enable efficient execution. Overall, {\toolName} provides a more favorable accuracy--efficiency trade-off than the alternatives: although it is slower than Marlin, it preserves higher accuracy and incurs less overhead than cuSPARSE.

\begin{table}[h]
\small
\caption{Throughput comparison in decoding (tokens/s).}
\label{tab:inference}
\resizebox{\linewidth}{!}{%
\begin{tabular}{cccccc}
   \toprule
Model                   & Batch Size & BF16             & Marlin              & cuSPARSE           & {\toolName}              \\ \midrule
\multirow{2}{*}{LLaMA3} & 8          & 56.9 & 138.5 & 90.7 & 114.4 \\
                        & 16         & 56.1 & 132.3  & 75.3 & 107.6   \\ \midrule
\multirow{2}{*}{Qwen2.5}   & 8          & OOM & 81.4 & 52.5 & 68.3 \\
                        & 16         & OOM & 79.4  & 41.0 & 64.2  \\    \bottomrule 
\end{tabular}

}
\end{table}

\section{Related Work}
Quantization~\cite{ashkboos_quarot_nodate,lin_awq_2025,dettmers_spqr_2023,dettmers_llmint8_nodate,frantar_gptq_2023,kim_squeezellm_2024} is widely used to improve the efficiency of LLM inference; however, outliers in Transformer weights remain a major obstacle to low-bit quantization. Prior work has mitigated weight outliers through value redistribution~\cite{lin_awq_2025,guo_olive_2023} or matrix-transformation-based methods~\cite{ashkboos_quarot_nodate,liu_spinquant_2025}. A complementary approach, \emph{outlier-aware quantization}~\cite{kim_squeezellm_2024,dettmers_llmint8_nodate,huang_slim-llm_2024,shang_pb_llm_2023}, retains a small subset of outlier weights in higher precision while quantizing the remaining weights to low-bit formats. Although this strategy improves the accuracy--compression trade-off, it introduces a mixed dense–sparse computation pattern during inference.

In practice, extracting and applying high-precision outliers typically requires an SpMM operator. Existing implementations~\cite{kim_squeezellm_2024,shang_pb_llm_2023,huang_slim-llm_2024} commonly rely on general-purpose sparse libraries such as cuSPARSE~\cite{cusparse} and Sputnik~\cite{gale_sparse_2020}. However, processing outliers through a separate sparse path from the quantized dense computation incurs redundant memory traffic and kernel launch overhead, reducing end-to-end throughput. To address this inefficiency, we propose a fused kernel that integrates quantized dense computation with outlier handling, streamlining the SpMM component and improving overall inference efficiency.

Our design is orthogonal to kernel optimizations for other quantization operators. Marlin~\cite{frantar2025marlin} implements a highly optimized dense low-bit GEMM kernel for weight-only quantization, but does not address the sparse outlier component introduced by outlier-aware weight quantization. MixQ~\cite{chen_mixq_nodate} targets activation outliers in weight--activation quantization, and FLUTE~\cite{guo_fast_2024} accelerates lookup-based quantization with specialized kernels. {\toolName} focuses on the mixed dense--sparse computation induced by outlier-aware weight quantization, which differs from the operator formulations targeted by these prior systems.

\section{Conclusion}
This paper presents \textbf{\toolName}, a unified execution framework for outlier-aware quantization that integrates low-bit GEMM with high-precision SpMM. By jointly leveraging sparse--dense tiling, a GPU-efficient outlier encoding format, and pipelined execution, {\toolName} reduces memory-access overhead while sustaining high throughput across diverse GPU architectures. Experiments show that {\toolName} achieves a $2.74\times$--$4.18\times$ speedup over cuBLAS BF16 and up to a $1.53\times$ improvement over the best existing outlier-aware implementation. Moreover, {\formatName} reduces storage and decoding overhead compared with CSR. In end-to-end quantized inference, {\toolName} delivers a $2.01\times$ speedup. Currently, {\toolName} targets INT4 weight-only quantization with BF16 outliers and relies on offline preprocessing to generate the Tile-COO representation; future work will extend the fused design to lower-precision formats such as INT2.

\bibliographystyle{ACM-Reference-Format}
\bibliography{ref}

\end{document}